\documentclass[12pt]{article}
\usepackage{amsmath,amssymb,exscale}
\usepackage{epsfig,psfrag}
\usepackage[usenames]{color}
\usepackage{amscd,shadow,fancybox,booktabs}
\input epsf

\newcommand{\be}{\begin{equation}}
\newcommand{\ee}{\end{equation}}
\newcommand{\bear}{\begin{eqnarray}}
\newcommand{\eear}{\end{eqnarray}}
\newcommand{\ba}{\begin{array}}
\newcommand{\ea}{\end{array}}

\newcommand{\gae}{\begin{array}{c}\,\sim\vspace{-21pt}\\>
\end{array}}

\title{
\begin{flushright}  
\normalsize{  
FERMILAB-PUB-09-641-T\\ 
}  
\end{flushright}  
\vspace{1.5cm}  
\Large \textbf{The Lepton Sector of a Fourth Generation}  
\\[0.5cm]}
\author{
\large  
G. Burdman$^{a,b}$, L. Da Rold$^c$, 
R. D. Matheus$^b$
\\ \\
[0.5cm]
$^a$\normalsize\emph{Fermi Nacional Accelerator Laboratory, Batavia, Illinois, 
USA}\\ 
$^b$\normalsize\emph{Instituto de F\'{i}sica, Universidade de S\~{a}o Paulo, 
S\~{a}o Paulo, Brazil}\\
$^c$\normalsize\emph{Centro At\'{o}mico Bariloche, Bariloche, Argentina}}

\date{}
\begin{document}  
\setcounter{page}{0}  
\maketitle  

\vspace*{0.5cm}  
\begin{abstract} 
\vspace*{0.5cm}
In extensions of the standard model with a heavy fourth generation 
one important question is what makes the fourth-generation lepton
sector, particularly the neutrinos, so  
different from the lighter three generations. 
We study this question in the 
context of models of electroweak symmetry breaking 
in warped extra dimensions, where the 
flavor hierarchy is generated by the localization of the zero-mode 
fermions in the extra dimension. In this setup the Higgs sector is localized 
near the infrared brane, 
whereas the Majorana mass term is localized at the ultraviolet brane. 
As a result, light neutrinos
are almost entirely Majorana particles, whereas the fourth generation neutrino 
is mostly a Dirac fermion. 
We show that it is possible to obtain heavy fourth-generation 
leptons in regions of parameter space where the light neutrino masses and 
mixings are compatible with observation. 
We study the impact of these bounds, as well as the ones from 
lepton flavor violation, on the phenomenology of these models.
\end{abstract}


\thispagestyle{empty}  
\newpage  
  
\setcounter{page}{1}  

\baselineskip18pt

\section{Introduction}
Perhaps the  simplest extension of the standard model is to 
allow for a chiral fourth generation. 
An obvious objection to this addition is the fact that  
the number of light neutrinos is 3, as accurately measured for instance
in $e^+e^-$ collisions~\cite{pdg}. Thus, the fourth-generation neutrino 
should be at least heavier than $M_Z/2$, which appears unnatural. 
Similarly, the fourth-generation charged lepton should have masses of the 
$O(100)$~GeV. It is possible to realize this situation
in theories with one compact extra dimension with an AdS metric~\cite{rs1}. 
In theories of electroweak symmetry breaking in these backgrounds the Higgs field 
must be localized close to the so-called infra-red (IR) brane in order to address the hierarchy problem.
The fermion mass hierarchy,  can then be naturally realized 
by localizing the zero-mode fermions close to or away from the IR brane. 
Then, if a fourth generation is added in this scenario, it suffices to localize its zero modes
close to the IR brane in order for them to have large enough masses.
This was the construction used in Ref.~\cite{bd} where, in addition, 
the resulting strong couplings of fourth-generation quarks to KK gluons were used to 
trigger electroweak symmetry breaking~\cite{bhl}. Since a heavy fourth generation 
must have large couplings to the dynamics responsible
for electroweak symmetry breaking, it is natural to consider it 
in association with strongly coupled TeV-scale physics. 
The phenomenology of the strongly coupled quark sector of a fourth generation was studied in this 
context in Ref.~\cite{d4}. 
However, it is interesting to 
study the lepton sector of a fourth generation in warped extra dimensions independently of 
the origin of the Higgs sector. Recently, the constraints on a standard model 
fourth generation have been re-examined~\cite{fourgen}, leading to the realization that 
a chiral fourth generation is allowed by electroweak precision constraints as long as 
the Higgs is heavier than in the standard model fits, and the mass differences 
in the isospin doublets are kept below $M_W$. Constraints might be even weaker in 
extensions of the standard model with a chiral fourth generation. 

In this paper we consider theories with one compact extra dimension with AdS metric (AdS$_5$) 
assuming the presence of four chiral generations in the bulk. For the most part, we will 
not need to assume a specific form of the Higgs sector, as long as the physical Higgs is 
localized close to the IR. Thus, most of our discussion applies to the generic model 
with an IR-localized Higgs, as well as to  composite Higgs models~\cite{so5} and the 
fourth-generation condensation scenario~\cite{bd}. Our goal is to show that 
the parameter space in the lepton sector of these models with a heavy 
chiral fourth generation  results in the correct pattern of neutrino masses and 
mixings, for acceptable values of the fourth-generation lepton masses and their mixing 
with the lighter generations. The resulting allowed parameter space in couplings and masses can 
then be used to study the phenomenology of these fourth generation leptons at colliders, particularly 
at the Large Hadron Collider (LHC).

In the next section we describe the 5D model of fourth generation leptons. 
We study the spectrum of the zero-mode leptons, as well as 
the constraints imposed by light neutrino masses and mixings, and by the mixing of 
the fourth generation to the lighter leptons. 
In Section~\ref{fcnc} we consider the constraints arising from flavor-changing processes, such
as $\mu^-\to e^-e^+e^-$ and $\mu^--e^-$ conversion in nuclei.
In Section~\ref{sec-embedding} we list the possible representations that 
leptons can take in the bulk gauge theory, which we take to be 
$SU(3)_c\times SU(2)_L\times SU(2)_R\times U(1)_X$~\cite{Agashe:2003zs}. 
This will be of use when studying the phenomenology of the lepton sector, which is done 
in Section~\ref{sec-phenomenology}, where we define the allowed parameter space and 
show that a typical occurrence in these models is the appearance of
light Kaluza-Klein (KK)  
modes of leptons. 
Here we also show the couplings of the zero-mode fourth-generation
leptons to gauge bosons, 
which will determine their collider phenomenology. We conclude in Section~\ref{conclusions}.

\section{Fourth Generation Leptons in AdS$_5$}
\label{model}
The model we consider here originates from a five-dimensional theory 
in an AdS background with the metric defined by~\cite{rs1} 
\be
ds^2 = e^{-2\sigma(y)}\,\eta_{\mu\nu}\,dx^\mu\,dx^\nu - dy^2~,
\label{metric}
\ee
where $\sigma(y)=k|y|$ and $k$ is the AdS curvature, $y$ is the coordinate in the extra dimension and it is bound to be in a segment  
between 0 (ultra-violet or UV) and L (IR).
The electroweak gauge group in the bulk is $SU(2)_L\times SU(2)_R\times U(1)_X$, with the hypercharge defined as $Y=T^3_R+X$, 
where $T^3_R$ is the $SU(2)_R$ isospin and $X$ is the charge under $U(1)_X$~\cite{Agashe:2003zs}. The bulk gauge symmetry is broken 
down to the SM gauge group by boundary conditions in the UV which lead to $SU(2)_R\times U(1)_X\to U(1)_Y$. 
However, as we will show below, many properties of the lepton model are independent of the choice of 
SU(2)$_R$ embedding for the fermions.

We consider four generations of leptons propagating in the 5D bulk: 
\begin{equation}
\xi^l_i \ , \quad 
\xi^\nu_i \ , \quad 
\xi^e_i \ ; \quad i=1,\dots 4.
\end{equation}
\noindent with corresponding bulk 5D masses $c_{l,\nu,e}^i$  in units of the AdS curvature $k$ .
The boundary conditions allow us to obtain a spectrum of zero modes that reproduces 
the SM lepton spectrum, plus a SM-like fourth generation with the SU(2)$_L$ doublets arising from $\xi^l$ and singlets from $\xi^{\nu,e}$. 

Since the $SU(2)_R\times U(1)_X$ gauge symmetry is broken to $U(1)_Y$ by boundary conditions in the UV,
a Majorana mass for bulk fermions transforming under $SU(2)_R$ can only be added on the UV boundary.  
So we introduce a UV-localized Majorana mass term for the right-handed neutrino of the form 
\begin{equation}\label{Majorana-mass}
\mathcal{S}_{0}=\int dx^4 dy\sqrt{g}\ \left(-\frac{1}{2}\,\bar\xi^{\nu c}_R M_{UV}\xi^\nu_R+\rm{h.c.} \right)\frac{\delta(y)}{\Lambda_{UV}}\ ,
\end{equation}
\noindent where $M_{UV}$ is a  $4\times4$ matrix, generation indexes are suppressed and $\Lambda_{UV}$ is the UV cutoff, typically the Planck 
mass $M_P$. 

We also consider Dirac masses for the leptons in the 5D bulk. These arise from their 5D Yukawa interactions with the 
Higgs doublet
\begin{equation}\label{Dirac-mass}
\mathcal{L}_{5}=-\lambda_\nu \bar\xi^l H \xi^\nu-\lambda_e \bar\xi^l H \xi^e+\rm{h.c.} \ ,
\end{equation}
\noindent 
with $\lambda_{\nu,e}$ being $4\times4$ matrices and the Higgs $H$ localized towards the IR. In the scenario of Ref.~\cite{bd} 
the Higgs arises from the condensation of the zero-mode quarks of the fourth generation, but our treatment of the fourth generation lepton 
sector is general, and will remain independent of the details of the Higgs sector unless specified. 

In general we will see that the zero-mode spectrum furnishes a
complete representation 
of the SM gauge symmetry instead of the full 
5D gauge symmetry, since in most cases SU(2)$_R$ is only broken in the UV boundary.  
The only exception is the case where the 5D field $\xi^\nu$ is a
singlet under all gauge interactions, thus allowing a Majorana term in
the IR. We will treat this case separately.
but in general our results will be independent of the SU(2)$_R$
embedding unless we state otherwise.   
 
The mixing of the zero-modes  with the KK modes is not important for
the discussion of the zero-mode spectrum, 
but its effects will be considered in
Section~\ref{sec-phenomenology}. 
After the Higgs acquires a vacuum expectation value (VEV), 
integrating over the extra dimension Eqs.~(\ref{Majorana-mass}) and~(\ref{Dirac-mass}) leads to
\begin{eqnarray}\label{mass}
\mathcal{L}_{\rm{mass}}&=&-\frac{1}{2}\bar\nu^c_R \ M_{RR}^\nu\ \nu_R
-\bar\nu_L M^\nu_{LR}\nu_R-\bar e_L M^e_{LR}e_R+\rm{h.c.} \ ,
\end{eqnarray}
\noindent where generation indexes  are understood, and for 
notational simplicity we have hidden the super-index $(0)$ labeling the zero-modes. 
The mass matrices $M_{RR}$ and $M^{e,\nu}_{LR}$ depend on the Higgs and fermionic zero-mode wave 
functions and will be computed in the next section. 

In order to obtain the observed spectrum and mixings, we have to select the 
localization of the zero-modes in the extra dimension. 
As usual in 5D theories, the localization of the zero-modes 
is determined by the bulk mass parameters  $c^i_{l,\nu,e}$ in such a way that, 
with ${\cal O}(1)$ variations in them, we can change the localization from the UV to the IR 
boundary~\cite{gn,gp}. 
Since left-handed leptons form a SU(2)$_L$ doublet, within each
generation 
$e_L$ and $\nu_L$ must have the 
same localization.
In addition, to obtain a pattern of neutrino mixings among the light 
generations compatible with observations, 
the overlap between the wave-functions of the left-handed zero-mode
leptons and 
the Higgs has to be of the same order 
for the first three generations. This can be done, for example, by
choosing localizations  
towards the IR, with $-1/2<c^i_l<1/2$. 
However, since  in the present model the fourth generation is also localized towards the IR, 
this scenario would lead to large mixings of the three light generations with the fourth generation. 
To avoid this, we will localize the left-handed fermions of the first
three generations 
near the UV boundary, $c^i_l\sim 0.6$. 
In this case the left-handed leptons have an exponentially suppressed overlap with the Higgs. Thus, to obtain same order Yukawas for the light 
generations, we impose that all the left-handed bulk leptons leading to the 
light generations have the same localization. This corresponds to 
\begin{equation}\label{sym-c}
c^i_l=c_l \ ,  \qquad i=1,2,3 \ .
\end{equation}
The requirements on the right-handed neutrinos of the first three
generations are that they have to be somewhat UV-localized but not too close to the UV so that their Majorana mass is not 
too large ($\sim M_P$) compared to the needed value for a successful see-saw mechanism. 
We will next discuss the specific values of 
$c_\nu^i$ that satisfy this constraint, but for now we can see that in order to obtain 
a single effective Majorana mass scale in the effective 4D theory of three generations of light neutrinos, we need 
to impose $c_\nu^i=c_\nu$ for $i=1,2,3$. These assumptions can be cast as the presence of a
global family symmetry in the 5D bulk involving the first three generations~\cite{perez,Csaki:2008qq}.
Finally, to obtain a heavy fourth generation of leptons  
we allow $l_L^4$ and $\nu_R^4$ to have different localizations from the ones chosen above for $i=1,2,3$, which as we show below must be 
toward the IR brane.
Throughout we consider anarchic Yukawas $\lambda_{\nu,e}$, meaning that all the entries of these matrices are of the same 
order: $\lambda_{\nu,e}^{ij}=\mathcal{O}(1)$, for $i,j=1,2,3,4$. 
Also for simplicity, we choose the UV-localized Majorana mass matrix as $M_{UV}=M_R\to M_R\bf{1}$, with $M_R$  a number of order $\sim M_{P}$.

\subsection{5D Lepton Embeddings}
\label{sec-embedding}
To complete the 5D model, we consider here  
the possible embedding of the lepton sector in the 5D bulk gauge
theory $SU(3)_c\times SU(2)_L\times SU(2)_R\times U(1)_X$.
As we will see in Section~\ref{sec-phenomenology}, some
phenomenological details depend on such embeddings. 
On the other hand, there are some generic features of the model which
are mostly independent of the chosen lepton 
representation in the 5D bulk. 
We start by pointing out that the Higgs is a $(\bf{2},\bf{2})_0$, with $\langle H\rangle \propto \bf{1}_{2\times 2}$. 
Therefore, the requirement that the mass term (\ref{Dirac-mass}) be a singlet, determines the transformation properties of the leptons.

\paragraph{\em Embedding 1}~\\
\label{sec-embedding1}
\noindent
A possible choice of bulk lepton representation under $SU(2)_L\times SU(2)_R\times U(1)_X$ can be 
\begin{equation}
\xi^l_i=({\bf{2}},{\bf{1}})_{-1/2}; \qquad 
\xi^\nu_i=({\bf 1},{\bf 2})_{-1/2}; \qquad 
\xi^e_i=({\bf{1}},{\bf{2}})_{-1/2}.
\label{emb1}
\end{equation}
The boundary conditions can be schematically written as
\begin{eqnarray}
\xi^{l}&=&
 \left[ \begin{array}{lr}
   \xi^l_L(++) & \xi^l_R(--)
 \end{array}\right]\ ,
\qquad
\xi^{\nu}=
 \left[ \begin{array}{cc}
   \xi^\nu_L=\left[\begin{array}{c}\nu'_L(--)\\ e'_L(+-)\end{array}\right]
&
   \xi^\nu_R=\left[\begin{array}{c}\nu_R(++)\\ e'_R(-+)\end{array}\right]
 \end{array}\right]\ , \nonumber
\\
\xi^{e}&=&
 \left[ \begin{array}{cc}
   \xi^e_L=\left[\begin{array}{c}\nu^{''}_L(+-)\\ e^{''}_L(--)\end{array}\right]
&
   \xi^e_R=\left[\begin{array}{c}\nu^{''}_R(-+)\\ e_R(++)\end{array}\right]
 \end{array}\right] \ ,
\end{eqnarray}
where $(++)$ refers to Newman boundary conditions on both the UV and the IR branes, $(--)$ to Dirichlet 
boundary conditions, etc.  
If we consider the Higgs localized in the IR, Eq.~(\ref{Dirac-mass}) leads to
\begin{equation}
{\cal L}_{\rm mass}=\bar \xi^l_L (m_\nu \xi^\nu_R+m_e \xi^e_R)|_{L_1}+{\rm h.c.}
= [\bar \nu_L (m_\nu \nu_R+m_e\nu^{''}_R)+\bar e_L (m_\nu e'_R+m_e e_R)]|_{L_1}+{\rm h.c.} \ ,
\end{equation}
with $\xi^{lt}_L=(\nu_L,e_L)$ and $m_{e,\nu}=\lambda_{e,\nu}v/\sqrt{2}$.

\paragraph{\em Embedding 2}~\\
Another possibility is 
\begin{eqnarray}
\xi^l&=&({\bf{2}},{\bf{2}})_{-1}=
 \left[ \begin{array}{cc}
   L_L(++) & L_R(--) \\ L'_L(-+) & L'_R(+-) 
  \end{array}\right]
 ; \\ 
\xi^\nu&=&({\bf 1},{\bf 3})_{-1}=
 \left[ \begin{array}{cc}
   \xi^\nu_L=\left[\begin{array}{c}\nu^{''}_L(--)\\ e^{'}_L(+-)\\ \chi_L(+-)\end{array}\right]
   &
   \xi^\nu_R=\left[\begin{array}{c}\nu_R(++)\\ e'_R(-+)\\ \chi_R(-+)\end{array}\right]
 \end{array}\right]
 ; \\ 
\xi^e&=&({\bf{1}},{\bf{1}})_{-1}=
\left[ \begin{array}{cc}
   e^{''}_L(--) & e_R(++)
 \end{array}\right]\ ,
\end{eqnarray}
\noindent where $L$ and $L'$ are the SU(2)$_L$ doublets with $T^{3R}=\pm 1/2$ respectively, contained in $\xi^l$. 
The boundary conditions for $L$ account for the zero-mode SM doublet, 
whereas $L'$ does not have zero-modes and gives rise to exotic KK leptons with electric charge $Q=-2$, as well as $\xi_\nu$. 
After EWSB the fermions with equal charges are mixed by the Higgs VEV.

\paragraph{\em Embedding 3}~\\
Similar to embedding 2, but:
\begin{eqnarray}
\xi^e&=&({\bf 1},{\bf 3})_{-1}=
 \left[ \begin{array}{cc}
   \xi^e_L=\left[\begin{array}{c}\nu^{'}_L(+-)\\ e^{''}_L(--)\\ \chi'_L(+-)\end{array}\right]
   &
   \xi^e_R=\left[\begin{array}{c}\nu'_R(-+)\\ e_R(++)\\ \chi'_R(-+)\end{array}\right]
 \end{array}\right]\ .
\end{eqnarray}

\paragraph{\em Embedding 4}
In this case the 5D leptons are:
\begin{eqnarray}
\xi^l&=&({\bf{2}},{\bf{2}})_{0}=
 \left[ \begin{array}{cc}
   L'_L(-+) & L'_R(+-) \\ L_L(++) & L_R(--) 
  \end{array}\right]
 ; \label{xil-model4}\\ 
\xi^\nu&=&({\bf 1},{\bf 3})_{0}=
 \left[ \begin{array}{cc}
   \xi^\nu_L=\left[\begin{array}{c}\chi_L(+-)\\ \nu^{''}_L(--)\\ e'_L(+-)\end{array}\right]
   &
   \xi^\nu_R=\left[\begin{array}{c}\chi_R(-+)\\ \nu_R(++)\\ e'_R(-+)\end{array}\right]
 \end{array}\right]
 ; \\ 
\xi^e&=&({\bf{1}},{\bf{3}})_{0}=
 \left[ \begin{array}{cc}
   \xi^e_L=\left[\begin{array}{c}\chi'_L(+-)\\ \nu^{'}_L(+-)\\ e^{''}_L(--)\end{array}\right]
   &
   \xi^e_R=\left[\begin{array}{c}\chi'_R(-+)\\ \nu'_R(-+)\\ e_R(++)\end{array}\right]
 \end{array}\right]\ ,
\end{eqnarray}
\noindent where $L'$ and $L$ are the SU(2)$_L$ doublets with $T^{3R}=\pm 1/2$ respectively, contained in $\xi^l$.

\paragraph{\em Embedding 5}~\\
Similar to embedding 4, but now $\xi^\nu$ is a 5D singlet.
\begin{equation}
\xi^\nu=({\bf{1}},{\bf{1}})_{0}=
\left[ \begin{array}{cc}
   \nu^{''}_L(--) & \nu_R(++)
 \end{array}\right]\ .
\label{xinu-model5}
\end{equation}

Unlike for the previous four embeddings, in this case it is possible
to have Majorana mass terms in locations other than the UV boundary,
since $\xi^\nu$ is a gauge singlet. In particular, it is possible to
write a Majorana mass term in the IR, $M_{IR}$. This will result in an 
order-one splitting of the masses of the Majorana components of the
fourth-generation zero-mode neutrinos. As a result, in this embedding
the fourth-generation zero-mode states are Majorana neutrinos, just as
the first three-generation neutrinos. We will consider their spectrum
in more detail below.

\subsection{Charged leptons}
\label{se-electrons}
Charged lepton masses are determined by the overlap of the left-handed and 
right-handed zero modes with the IR-localized Higgs.  
Their mass matrix is given by
\begin{eqnarray}\label{Me}
M_{LR}^{e,ij}=m_1 \lambda_e^{ij}f(c^i_l)f(-c^j_e)\alpha(c^i_l,-c^j_e) \ , \\
f(c)=\left[\frac{1-2c}{1-x^{1-2c}} \right]^{1/2} \ , \quad x=e^{-k\,L} \ ,
\end{eqnarray}
\noindent 
where the the functions $\alpha$ and $m_1$ describe the Higgs localization and the Higgs VEV, respectively. 
For instance, if the Higgs arises from the condensation of the zero modes of the fourth-generation quarks, $m_1$ and $\alpha$ are defined 
by~\cite{bd}
\begin{eqnarray}\label{m-condensation-q}
m_1&=&\frac{\langle\bar u^{4(0)}_Lu^{4(0)}_R\rangle}{k^2\,x^2}\,
\left(\frac{k}{M_P}\right)^3 \ ,\\
~&~&~\nonumber\\
\alpha(c_i,-c_j)&=&\frac{f(c^4_q)f(-c^4_u)(1-x^{4-c_q+c_u-c_i+c_j})}{(4-c_q+c_u-c_i+c_j)} \ .
\label{a-condensation-q}
\end{eqnarray}
\noindent On the other hand, if the Higgs arises from the zero-mode of a fundamental 5D scalar field we have
\begin{eqnarray}\label{m-condensation-h}
m_1&=&\langle H\rangle  \ ,
\\
\alpha(c_i,-c_j)&=&\frac{(1-x^{2+\beta-c_i+c_j})\sqrt{2(1+\beta)}}{(2+\beta-c_i+c_j)(1-x^{2+2\beta})^{1/2}} \ ,
\label{a-condensation-h}
\end{eqnarray}
\noindent with $\beta$ a function of the the 5D Higgs mass: $\beta=\sqrt{4+m_H^2L^2}$, (see for example Ref.~\cite{Cacciapaglia:2006mz}). 

The masses of the charged leptons depend on the localization of the right-handed 
component, $e^i_R$, that is controlled by the bulk mass parameter $c^i_e$. 
The left-handed components of the light generations are almost delocalized, 
in such a way that we can obtain a light electron (a heavy $\tau^-$) by 
localizing its right-handed component towards the UV (IR). 
To obtain a heavy fourth generation lepton both chiralities of $e^4$ must be
 localized towards the IR. 
The charged-lepton mass matrix $M^e_{LR}$ is diagonalized as usual, by
left and right unitary transformations $A^e_{L,R}$, such that the 
diagonal mass matrix is 
\begin{equation}
M^e_D= U^{L\dagger} M^e_{LR} U^R \ .
\end{equation}

The size of the mixings among the zero modes determines the size of the flavor-violating processes. 
Using Eq.~(\ref{Me}) we expect the mixings to be of order~\cite{Agashe:2004cp,Casagrande:2008hr,Bauer:2009cf}
\begin{equation}
U^L_{ij}\sim\frac{{\rm min}[f(c^i_l),f(c^j_l)]}{{\rm max}[f(c^i_l),f(c^j_l)]}\ , \qquad
U^R_{ij}\sim\frac{{\rm min}[f(-c^i_e),f(-c^j_e)]}{{\rm max}[f(-c^i_e),f(-c^j_e)]} \ ,
\end{equation}
\noindent In the scenario with $c^i_l\simeq c_l$, since the Yukawas are all 
of the same order, there are no suppression factors in the mixings 
between the left-handed components of the light generations: 
\begin{equation}
U^L_{ij}\sim{\cal O}(1) \ , \qquad i,j=1,2,3.
\end{equation}
\noindent In fact, we have enforced this result by adjusting the localization of 
left-handed components of the light generations. On the other hand, using Eq.~(\ref{Me}) 
and $f(c^i_l)\simeq f(c^j_l)$, we expect the mixings between the right-handed components of the 
light generations to be of order
\begin{equation}
U^R_{ij}\sim\frac{{\rm min}[m_i,m_j]}{{\rm max}[m_i,m_j]} \ , \qquad m_{i,j}=m_e,m_\mu,m_\tau.
\end{equation}
\noindent Thus we obtain a hierarchical $U^R$ with small mixings between the light generations.

Since both the left- and right-handed components of $e^4$ are localized towards the IR, the mixings between the light generations and $e^4$ are of order:
\begin{equation}\label{mixi4}
U^L_{i4}\sim\frac{f(c^i_l)}{f(c^4_l)} \ , \qquad  U^R_{i4}\sim\frac{f(-c^i_e)}{f(-c^4_e)} \ , \qquad i=1,2,3.
\end{equation}

In our model the $\tau_R$ has to be almost delocalized in order to obtain the correct value of $m_\tau$. Therefore, the mixing between $\tau_R$ and $e^4_R$ can 
be rather large, as can be seen from Eq.~(\ref{mixi4}). In our numerical scan (see next sections) we obtained
\begin{equation}\label{mixtau4}
U^L_{\tau4}\sim\frac{1}{2}\sqrt{\frac{m_\tau}{m_{e^4}}} \ , \qquad  U^R_{\tau 4}\sim 2\sqrt{\frac{m_\tau}{m_{e^4}}} \ .
\end{equation}
\noindent The large mixing $U^R_{\tau 4}$ arises as a consequence of the rather large $\tau$-mass and the partial UV localization of $\tau_L$. 
This result has important phenomenological consequences for $e^4$ FCNC decays, as we will discuss in Section~\ref{subsec:couplings}.

\subsection{Neutrinos}
\label{sec-neutrinos}
In what follows we consider the neutrino spectrum for embeddings 1 to
4. The case of embedding 5 will be consdered separately. 
There are two mass terms for the neutrinos: a Majorana mass for the 
right-handed components, which will affect almost exclusively the ones localized in the UV; 
and a Dirac mass determined by the overlap with the IR-localized Higgs,
as seen in Eq.~(\ref{mass}).
The Dirac mass is given by Eq.~(\ref{Me}), changing the index $e\to \nu$. 
On the other hand, the effective Majorana mass matrix for the zero-mode right-handed 
neutrinos can be written as
\begin{eqnarray}\label{MR}
M_{R}^{\nu,ij}\simeq O(1)\,k\,F(-c_i)F(-c_j) \ , \quad
F(c)=\left[\frac{2c-1}{1-x^{2c-1}} \right]^{1/2} \ .
\end{eqnarray}
\noindent 
The spectrum of the light SM neutrinos is generated by the usual see-saw mechanism. 
As discussed above, the left-handed components of the first three generations  
are almost delocalized. 
The right-handed 
components have bulk mass parameters in the range
\begin{equation}
-0.4\lesssim c^i_\nu\lesssim -0.28 \qquad i=1,2,3 \ .
\end{equation}
leading also to wave functions in the bulk, although somewhat more localized towards the UV. 

\begin{figure}[h]
\centering
\psfrag{xAxis}{$c_{\nu}^4$}
\psfrag{yAxis}{$m_{L^4}, m_{R^4}$ (TeV)}
\includegraphics[width=.6\textwidth]{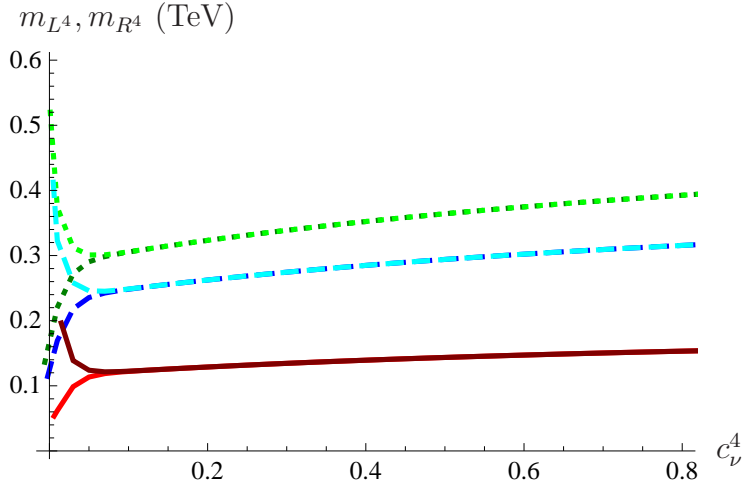}
\psfrag{cn4}{$c_\nu^4$}
\caption{\em Mass of the heavy fourth generation Majorana neutrinos, as a function of $c_\nu^4$. The different lines correspond 
to different values of $c_l^4=-0.4,0,0.4$. For $c_\nu^4$ fixed, the larger $m_{\nu^4}$ corresponds to the smaller $c_l^4$, 
since in this case $\nu^4_L$ is more localized in the IR. As long as $\nu^4_R$ is near the UV, the Majorana mass becomes large, 
and the see-saw mechanism starts working, splitting the neutrinos in two: one very light and another very heavy, this happens 
for $c_\nu^4\lesssim 0$.}
\label{Fig-mass-n4}
\end{figure}

Since we want a heavy fourth generation neutrino, we have to localize $\nu^4_R$ towards the IR,
in order to avoid the effect of the UV-localized  
Majorana mass, which would result in an unwanted see-saw and a light $\nu^4$. 
Due to the exponential localization 
of the right-handed neutrino, the mass of $\nu^4$ has a strong dependence on  
$c_\nu^4$ (the left-handed component 
is localized in the IR to obtain a heavy $e^4$). 
In Fig.~\ref{Fig-mass-n4} we show $m_{\nu^4}$ as a function of $c_\nu^4$, 
for different values of $c_l^4=-0.4,0,0.4$. We have fixed $M_{UV}=0.3\,k$, all the Yukawas are equal to unity, 
and we have neglected the mixings between generations for simplicity.~\footnote{For a heavy neutrino, the mixings are very small, 
thus we do not expect important corrections in Fig.~\ref{Fig-mass-n4} from their effects.} For $c_\nu^4\lesssim 0$, $\nu^4_R$ 
has a sizable overlap with the UV and the see-saw mechanism becomes efficient, resulting in two Majorana neutrinos: one that is very light, and another 
one very heavy. For $c_\nu^4\gtrsim 0$ the Majorana term becomes small enough and the Dirac mass dominates, resulting in 
both Majorana neutrinos being almost degenerate, or effectively in a Dirac fermion $\nu^4$.

The right-handed neutrinos of the first three generations 
$\nu^{e,\mu,\tau}_R$, being localized towards the UV, have a very large effective Majorana mass ($\sim 10^{-5} M_{Pl}$, 
depending on $c_\nu$). For this reason we integrate them out using the tree level equations of motion. The effective theory contains five 
Majorana neutrinos, three very light that reproduce the SM spectrum, and two almost degenerate heavy neutrinos which constitute an effective 
Dirac fourth-generation neutrino, 
as shown in Figure~\ref{Fig-mass-n4}. The mass term of the effective theory can be written as
\begin{equation}\label{Meffnu}
\mathcal{L}_{\rm{eff}}=-\frac{1}{2}\bar N M_{\rm{eff}} N^c + \rm{h.c.}
\end{equation}
\noindent where $N$ is defined by
\begin{eqnarray}
N^t=(\nu^e_L,\nu^\mu_L,\nu^\tau_L,\nu^4_L,\nu^{4\ c}_R)\ ,
\end{eqnarray}
\noindent and $M_{\rm{eff}}$ is the effective mass matrix, that can be written in terms of the elements of 
$M^\nu_{RR}$ and $M^\nu_{LR}$ defined in Eq.~(\ref{mass}), as shown in Appendix~\ref{Ap-diagonalization}. 
The neutrino spectrum is given by the eigenvalues of $M_{\rm{eff}}$ and the neutrino mass-eigenstates by the eigenvectors of $M_{\rm{eff}}$. 
Thus we diagonalize $M_{\rm eff}$ with an orthogonal transformation $U^\nu$ defined by
\begin{equation}
M^\nu_D=U^{\nu t} M_{\rm eff}U^\nu \ .
\end{equation}
\noindent Expanding in inverse powers of the large Majorana mass $M^{\nu,ij}_R$, with $i,j=1,2,3$, 
the eigenvalues and eigenvectors corresponding to the heavy Majorana neutrinos, with masses $\sim{\cal O}(300)$ GeV, are given by:
\begin{eqnarray}\label{m45}
m^\nu_{4,5}&=&(\sum_{i=1}^4 M^{\nu,i4}_{LR}M^{\nu,4i}_{RL})^{1/2} \ , \\
\tilde{N}_{4,5}&=&(M^{\nu,14}_{LR},M^{\nu,24}_{LR},M^{\nu,34}_{LR},M^{\nu,44}_{LR},\pm m^\nu_{4,5})\frac{1}{m^\nu_{4,5}\sqrt{2}}\ ,
\label{v45}
\end{eqnarray}
Thus, at leading order the heavy Majorana neutrinos 
$\tilde{N}_4$ and $\tilde{N}_5$ are degenerate, as advertised earlier and shown in Figure~\ref{Fig-mass-n4}. 
It is possible to obtain similar expressions for the light eigenvectors and eigenvalues. 
We show these results in Appendix~\ref{Ap-diagonalization}.

The wealth of neutrino data collected in the last decade results in important constraints on 
these models through the precise measurements of neutrino mass differences and mixing angles. 
We now explore the possibility of reproducing  these data 
in the model described above, with anarchic Yukawa couplings 
\footnote{
However, we want to stress that this is not the full anarchic approach, since we have imposed a partial global flavor 
symmetry relating some of the 5D leptonic masses, as shown in  Eq.~(\ref{sym-c}). 
Had we considered anarchic 5D masses, we would have obtained hierarchical mixing, in contradiction with observation.}.
In the present scenario, considering only zero-mode fermions and integrating out the heavy 
right-handed neutrinos of the first three generations, we are left  with four charged leptons and five Majorana neutrinos, 
as mentioned earlier. 
Then the  charged current interactions in the physical basis are given by
\begin{equation}\label{LW}
{\cal L}_{CC}=\sum_{\alpha,a}\sum_n \frac{g_n}{\sqrt{2}}W_\mu^{+(n)}\bar\nu^a_L \gamma^\mu V_{a\alpha}e^\alpha_L+ {\rm h.c.} \ ,
\end{equation}
where $W^{(n)}$ and $g_n$ are the KK $W$'s and their couplings respectively, with $g_0=g$ and $g_n/g_0$ depending on the localization of the zero-mode leptons:
for the light (heavy) generations $g_1/g_0\sim{\cal O}(0.1)$ ($\sim\sqrt{2k\pi r}\sim{\cal O}(10)$). $V$ is the $5\times 4$ mixing matrix that can be 
expressed in terms of the neutral and charged rotation matrices $U^{\nu}$ and $U^{L}$:
\begin{equation}\label{V}
V_{a\alpha}=\sum_{i=1}^4 (U^{\nu t})_{ai}U^{L}_{i\alpha} \ , \qquad a=1,\dots 5 \ , \ \ \alpha=1,\dots 4 \ .
\end{equation}
The rotation matrix $U^\nu_{5\times5}$ is orthonormal, whereas $U^{L}_{4\times4}$ is unitary.
The mixing matrix $V$ satisfies the following constraints
\begin{equation}\label{unitarity}
(V^\dagger V)_{\alpha\beta}=\delta_{\alpha\beta} \ , \qquad (V\ V^\dagger)_{ab}=\delta_{ab}-(U^{\nu t})_{a5}U^{\nu}_{5b} \ .
\end{equation}
Using the results of Eqs.~(\ref{m45}) and~(\ref{v45}) one can show that, to leading order, Eq.~(\ref{unitarity}) reduces to:
\begin{equation}\label{unitarity2}
(V\ V^\dagger)=
\left[\begin{array}{cc}
1_{3\times3}&0\\
0&\left[\begin{array}{cc}1/2&1/2\\1/2&1/2\end{array}\right]
\end{array}\right]
\end{equation}
\noindent
If we consider only the first three generations, we can impose that  $V$ reproduces the observed neutrino mixing matrix. 
On the other hand, the mixings between  zero-modes of the fourth generation and the light leptons of the first three generations 
are mostly determined by the 
localization of the left-handed wave functions. They are experimentally constrained, mostly from the decay 
$\mu^-\to e^-\gamma$, which receives new contributions induced by the presence of the new neutrinos running in the loop, as well as  by the lack of unitarity of the $3\times 3$ mixing matrix. 
The contribution of a neutrino, with Dirac and right-handed Majorana masses, to the decay $\mu^-\to e^-\gamma$ can be written 
as~\cite{Cheng:1980tp}
\begin{eqnarray}
T_i&=&V^\dagger_{\mu i}V_{ie} H(m_i^2/m_W^2) \\
H(x)&=&\frac{1}{3(x-1)^4}(10-43x+78x^2-49x^3+4x^4+18x^3\log x) \ ,
\end{eqnarray}
leading to a branching fraction 
\begin{equation}\label{br}
BR(\mu\to e\gamma)=\frac{\Gamma[\mu\to e\gamma]}{\Gamma[\mu\to e\nu\bar\nu]}=\frac{3\alpha}{32\pi}|\sum_i T_i|^2 \ .
\end{equation}
In the present model we have to sum over $i=1, \dots 5$, with $m_i\ll m_w$ for $i=1,\dots 3$.  We obtain
\begin{eqnarray}
\sum_{i=1}^5 T_i\simeq \sum_{i=1,23}V^\dagger_{\mu i}V_{ie}\left[H(0)+H'(0)\frac{m_i^2}{m_W^2}\right]+
\sum_{i=4,5} V^\dagger_{\mu i}V_{ie}H\left(\frac{m_i^2}{m_W^2}\right) \ .
\end{eqnarray}
In the SM only the first sum is present, and the GIM mechanism cancels the first term in the square brackets: 
$\sum_{i=1}^3V^\dagger_{\mu i}V_{ie}=0$. In the present case we have instead
\begin{equation}
\sum_{i=1,23}V^\dagger_{\mu i}V_{ie}=-\sum_{i=4,5} V^\dagger_{\mu i}V_{ie} \ .
\end{equation}
which leads to 
\begin{eqnarray}\label{sumTimodel}
\sum_{i=1}^5 T_i\simeq \sum_{i=4,5} V^\dagger_{\mu i}V_{ie}\left[H\left(\frac{m_i^2}{m_W^2}\right)-H(0)\right]+ \dots \ ,
\end{eqnarray}
where the dots stand for a SM type contribution, which can be neglected since is very small compared to the current bounds.
Inserting Eq.~(\ref{sumTimodel}) into~(\ref{br}) we obtain an upper bound for $|V_{4\alpha}|$ and $|V_{5\alpha}|$ in the present model.
Taking as a reference value~\cite{pdg} $BR(\mu^-\to e^-\gamma)\lesssim 10^{-11}$ we obtain 
\begin{equation}\label{boundmuegamma}
V_{\mu,i}\sim V_{e,i}\lesssim 10^{-2}, \qquad i=4,5.
\end{equation}
In the next section we will consider additional constraints from FCNC effects.

\subsection{Model Parameter Scan}
\label{scan}
We have scanned over the model parameter space in order to find solutions compatible with all the available constraints.
In our scan we have  considered the following range for the 5D parameters: 
\begin{itemize}
\item 5D Yukawa couplings of order ${\cal O}(1)$: $-2\leq\lambda_{e,\nu,R}^{ij}\leq2$,\\ 
assuming real Yukawas for simplicity, 
\item 5D bulk mass parameters:\\ 
\begin{eqnarray}
0.6\leq c_l\leq 0.65,& -0.4\leq c_\nu\leq -0.28,& 0.4\leq c_e^i\leq 0.72 \nonumber\\
-0.5< c_l^4<0.5, & 0\leq c_\nu^4<0.5, & -0.5<c_e^4<0.5\nonumber
\end{eqnarray}
\end{itemize}
Finally, we considered $k\,e^{-kL}=1$~TeV (except in the cases where we explicitly mention a larger KK scale), 
and assumed a Higgs arising from the condensation of the quarks of the
fourth generation, 
see Eqs.~(\ref{m-condensation-q}) and~(\ref{a-condensation-q}), 
with $c_q^4<0$ and $c_u^4>0$. 
We observe that in this region of parameter space, it is possible to obtain lepton masses and mixings 
between the light states compatible with experiment. The same
solutions result in fourth 
generation-leptons heavy enough to have evaded direct detection 
bounds. 
We have also studied the amount of tuning in the space
$\{c^i_l,c^i_\nu\}, \ i=1,2,3$, 
needed to obtain the right pattern of mixings. 
We have allowed a small random variation of the 5D masses in our
numerical scan:  
$\{c^i_l=c_l+\Delta^i_l,c^i_\nu=c_\nu+\Delta^i_\nu\}$. 
Our results show that, in order to satisfy the constraints, $\Delta$
can not be larger 
than $\Delta^i_l\sim\Delta^i_\nu\sim 0.01$. 
As a benchmark point in the parameter space of 5D masses we consider
\begin{eqnarray}\label{becnchmark1}
c_l&=&0.59 \ , \qquad c_\nu=-0.34 \ , \qquad c_e=-0.73 \ , \qquad c_\mu=-0.61 \ , \qquad c_\tau=-0.52 \ , \nonumber \\
c_l^4&=&0 \ , \qquad c_\nu^4=0.30 \ , \qquad c_e^4=0.35 \ .
\end{eqnarray}
In Appendix~\ref{Ap-numbers} we show an example of a specific solution satisfying all the constraints.

\subsection{Neutrino Spectrum in Embedding~5}
\label{nuspecine5}
We discuss here the spectrum of zero-mode neutrinos in
Embedding~5, presented in Section~\ref{sec-embedding}. In it the
right-handed neutrino zero-modes come from $\xi^\nu$ which transforms
as $({\bf 1},{\bf 1})_0$ under $SU(2)_L\times SU(2)_R\times
U(1)_X$. This choice allows the presence of a Majorana mass term in
the IR brane 
\be
\mathcal{S}_{IR}=\int dx^4 dy\sqrt{g}\
\left(-\frac{1}{2}\,\bar\xi^{\nu c}_R 
{\cal M}_{IR}\xi^\nu_R+\rm{h.c.} \right)\frac{\delta(y-L)}{\Lambda_{UV}}\ ,
\ee
which induces a Majorana mass 
\be
M_{IR}^{ij}=O(1)\,(k e^{-k L})\,f(-c_i) f(-c_j)
\ee
for the 
zero-mode neutrinos.
Since the first three-generation neutrinos are UV-localized they will
not be significanty affected by $M_{IR}$. Thus the effect of $M_{IR}$
can be schematically described by 
\be
{\cal L} = -\frac{1}{2}\,\bar\nu^4_R M_{IR} \nu^{4 c}_R - \bar\nu_L^4
M^4_{LR}\nu^4_R + h.c.
\ee
where $M^4_{LR}$ is the zero-mode Dirac mass for the fourth-generation
neutrino, and we have neglected mixing with the lighter generations.
This leads to a  splitting between the two Majorana states induced by
$M_{IR}$, which is typically of order one. 
The light state is an admixture of $\nu^4_L$ and
$\nu^4_R$, but it has typically a larger fraction of $\nu^4_L$ as long
as $M_{IR}> M_{LR}^4$. The lighter state could be rather light given
that there is a mild see-saw controlled by 
${\cal M}_{IR}$. For instance, it is quite natural in this embedding
to consider a lighter Majorana states with masses as small as
$100~$GeV.

\section{Constraints from Flavor Violation}
\label{fcnc}
In this section we study the constraints on these models from lepton flavor violation. 
The bounds from the dipole operator inducing 
$\mu^-\to e^-\gamma$ were shown in the previous section. 
Here we concentrate on flavor violating decays of the charged leptons, 
such as  $\mu^-\to e^-e^+e^-$ and $\mu^--e^-$ conversion in nuclei. As we show below, there is also a 
significant contribution to the decay of $e^4$ through FCNC, resulting in  
$e^4\to\tau^- f\bar f$, with  $f$ a SM light fermion.

We consider the effects from the interactions between the would-be zero-mode leptons and 
the neutral vector bosons, neglecting the effects from Higgs boson exchange
since they are suppressed by the small SM lepton masses\footnote{See Refs.~\cite{Agashe:2004cp,Casagrande:2008hr,Bauer:2009cf,Agashe:2009di} for FCNC effects arising from the Higgs in composite 
Higgs models.}. Although the couplings between the KK vector bosons and the leptons are 
not universal due to the generation-dependent localization of the leptons, the neutral-current interactions are diagonal in the flavor basis before 
electroweak symmetry breaking (EWSB)
\begin{equation}\label{Zintdiagonal}
{\cal L}_{NC}=\sum_{a}\sum_n Z_\mu^{(n)}\bar e^a \gamma^\mu (g^{L}_{na}P_L+g^{R}_{na}P_R)e^a \ ,
\end{equation}
\noindent where $g^{L,R}_{0a}=g^{L,R}_{SM}$ is the usual SM Z-coupling. 
After EWSB, the Z boson arises primarily from the mixing between the zero and the first KK neutral modes,  $Z^{(0)}$ and $Z^{(1)}$ 
(we neglect the mixing with the heavier KK-modes for simplicity in this analysis). Diagonalizing the vector boson mass matrix, the physical Z boson at leading order 
is given by
\begin{equation}\label{z}
Z=Z^{(0)}-f\frac{m_Z^2}{m_{KK}^2}Z^{(1)} \ ,
\end{equation}
where $f\sim \sqrt{2\,k L}\sim{\cal O}(10)$ is a factor parametrizing the mixing, and in general has a mild dependence on the Higgs localization. 
Thus, since the Z boson has a sizable projection over the first KK mode, the neutral lepton interactions are not universal after EWSB. 
In order to obtain the neutral interactions for the mass eigenstates we rotate the flavor basis in (\ref{Zintdiagonal}), which results in
\begin{equation}\label{Zintmixing}
{\cal L}_{NC}=\sum_{a,b}\left[Z_\mu\bar e^a \gamma^\mu (g^{L}_{ab}P_L+g^{R}_{ab}P_R)e^b+\sum_n Z^n_\mu\bar e^a \gamma^\mu (g^{L}_{nab}P_L+g^{R}_{nab}P_R)e^b \right],
\end{equation}
\noindent where $Z^n$ stands for the heavy neutral vectors arising from the mixed KK modes. 
The flavor-violating Z couplings in the mass eigenbasis are given by
\begin{equation}\label{gZab}
g^{L,R}_{ab}=-f\frac{m_Z^2}{m_{KK}^2} (U^{L,R\dagger}G^{L,R}U^{L,R})_{ab}\ , 
\qquad G^{L,R}={\rm diag}(g^{L,R}_{1e},g^{L,R}_{1\mu},g^{L,R}_{1\tau},g^{L,R}_{14}) \ ,
\end{equation}
\noindent where $g^{L,R}_{1a}$ are the diagonal flavor dependent couplings with $Z^{(1)}$, defined in Eq.~(\ref{Zintdiagonal}). 

The flavor-violating couplings relevant for the $\mu^--e^-$ transitions are determined by
\begin{eqnarray}\label{gZviolating}
(U^{L,R\dagger}G^{L,R}U^{L,R})_{e\mu}=& &
U^{L,R\dagger}_{12}(g^{L,R}_{1\mu}-g^{L,R}_{1e})U^{L,R}_{22}+
U^{L,R\dagger}_{13}(g^{L,R}_{1\tau}-g^{L,R}_{1e})U^{L,R}_{32}+\nonumber \\& &
U^{L,R\dagger}_{14}(g^{L,R}_{1e^4}-g^{L,R}_{1e})U^{L,R}_{42} \ ,
\end{eqnarray}
\noindent where we have used the unitarity of the rotation matrices. 
There are similar expressions for the other flavor-violating neutral interactions $(U^{L,R\dagger}G^{L,R}U^{L,R})_{\alpha\beta}$, $\alpha\neq\beta$. 

Using the estimates of Section~\ref{se-electrons} we can obtain the size of the flavor-violating couplings. 
From (\ref{gZviolating}) we see that for the choice 
$c^i_l=c_l, \ i=1,2,3$, the couplings $g^L_{1i}$ have the same value for the light generations and the only contributions to flavor-violating processes 
in the left-handed sector are due to the fourth generation. 
Allowing for  small departures from universal localization, $c^i_l=c_l+\Delta^i_l$, we also obtain contributions from the light generations. 
Using the results from our numerical scan allowing $\Delta^i_l\sim 0.01$, the contributions to FCNC processes relevant for $\mu^--e^-$ transitions  come from the 
flavor-violating factors
\begin{eqnarray}\label{gZviolatingmuL}
U^{L\dagger}_{12}(g^{L}_{1\mu}-g^{L}_{1e})U^{L}_{22}&\sim&\frac{1}{2}\times 10^{-3}g^L\times \frac{1}{2} \sim 10^{-4}g^L \ ,\\
\label{gZviolatingtauL}
U^{L\dagger}_{13}(g^{L}_{1\tau}-g^{L}_{1e})U^{L}_{32}&\sim&\frac{1}{2}\times 10^{-3}g^L\times\frac{1}{2} \sim 10^{-4}g^L \ ,\\
\label{gZviolating4L} 
U^{L\dagger}_{14}(g^{L}_{1e^4}-g^{L}_{1e})U^{L}_{42}&\sim&10^{-2}\times 5g^L\times 10^{-2} \sim 5\times 10^{-4}g^L \ .
\end{eqnarray}
\noindent 
In (\ref{gZviolating4L}) we have considered the fact that $e^4_L$ is not completely localized in the IR, and for this reason the coupling with the KK vector 
is somewhat smaller than $f$, $g^L_{14}\sim f/2\sim 5$. On the other hand, since the light generations are localized towards the UV, 
their couplings to the KK vectors are suppressed by: $g^L_{1i}\sim1/f\sim{\cal O}(0.1)$ for $i=1,2,3$.
The dominance of the $e^4$ contribution suggested by Eqs.~(\ref{gZviolatingmuL}-\ref{gZviolating4L}) is confirmed in 
our more detailed numerical studies. 

Similarly, we can obtain  contributions of the right-handed sector, given by 
\begin{eqnarray}\label{gZviolatingmuR}
U^{R\dagger}_{12}(g^{R}_{1\mu}-g^{R}_{1e})U^{R}_{22}&\sim&\frac{m_e}{m_\mu}\times 10^{-3}g^R\times 1 \sim 10^{-5}g^R \ ,\\
\label{gZviolatingtauR}
U^{R\dagger}_{13}(g^{R}_{1\tau}-g^{R}_{1e})U^{R}_{32}&\sim&\frac{m_e}{m_\tau}\times 0.1g^R\times\frac{m_\mu}{m_\tau} \sim 10^{-5} g^R\ ,\\
\label{gZviolating4R}
U^{R\dagger}_{14}(g^{R}_{1e^4}-g^{R}_{1e})U^{R}_{42}&\sim&\frac{m_e}{m_{e^4}}\times 5g^R\times \frac{m_\mu}{m_{e^4}} \sim 5\times 10^{-5} g^R\ .
\end{eqnarray}
\noindent 
Again $e^4$ gives the dominant contribution. We also note that the flavor-violation effect in the left-handed  sector is an order of magnitude larger than the one coming from the right-handed sector. 
Taking into account all ${\cal O}(1)$ coefficients, a  numerical 
scan results in $(U^{R\dagger}G^{R}U^{R})_{e\mu}\sim1/3\times10^{-4}$.

The most stringent constraints come from the upper limits for $\mu^-\to e^-e^+e^-$ decay branching ratio, and from the $\mu^--e^-$ conversion rate in nuclei. Following the analysis of~\cite{Chang:2005ag} and \cite{Agashe:2006iy}, we  
define the relevant effective couplings by
\begin{eqnarray}\label{LFCNCeff}
-{\cal L}_{eff}=\frac{4G_F}{\sqrt{2}}&[&\!\!\!\! g_3(\bar e_R\gamma_\mu\mu_R)(\bar e_R\gamma_\mu e_R)+g_4(\bar e_L\gamma_\mu\mu_L)(\bar e_L\gamma_\mu e_L) \nonumber \\ 
&+&\!\!g_5(\bar e_R\gamma_\mu\mu_R)(\bar e_L\gamma_\mu e_L)+g_6(\bar e_L\gamma_\mu\mu_L)(\bar e_R\gamma_\mu e_R) \ ]+{\rm h.c.}\ .
\end{eqnarray}
\noindent 
with $G_F$ the Fermi constant. Normalizing to the $\mu^-\to e^-\bar\nu\nu$ branching ratio, 
we can write the branching ratio for $\mu^-\to e^-e^+e^-$ as
\begin{equation}
\label{mu-eee}
\frac{BR(\mu^-\to e^-e^+e^-)}{BR(\mu^-\to e^-\bar\nu\nu)}=2(|g_3|^2+|g_4|^2)+|g_5|^2+|g_6|^2 \ .
\end{equation}
\noindent 
The current experimental limit is $BR(\mu^-\to e^-e^+e^-)<10^{-12}$~\cite{pdg}. 
The $\mu^--e^-$ conversion rate in nuclei is given by~\cite{Agashe:2006iy,Kuno:1999jp}:
\begin{equation}\label{mu-e}
B_{\rm conv}(\mu-e)=\frac{2p_eE_eG_F^2m^2_\mu\alpha^3Z_{eff}^2Q_N^2}{\pi^2Z\Gamma_{\rm capt}}(|g^L_{e\mu}|^2+|g^R_{e\mu}|^2) \ ,
\end{equation}
\noindent 
where $g^{L,R}_{e\mu}$ are the flavor-violating Z couplings defined in 
(\ref{gZab}) and $\alpha$ is the fine structure constant. 
The other factors depend on the specific nuclei involved in the reaction and can 
be found in Ref.~\cite{Kuno:1999jp}. 
The most constraining limits arise from the SINDRUM II experiment at PSI using Au. The corresponding bound is
$B_{\rm conv}(\mu^--e^-)<0.7\times 10^{-12}$, at 90\% C.L.~\cite{pdg}.

In order to obtain the model predictions for the conversion rate, we consider  
both the contributions from the $Z$ as well as from the 
neutral KK vectors in (\ref{Zintmixing}). 
From (\ref{gZab}) the dominant contributions to the effective couplings 
$g_{3-6}$ are given by
\begin{eqnarray}\label{matchg3-6}
g_3\simeq \frac{g^{R}}{m_{KK^2}}\left(f-\frac{g^{R}_{1e}}{g^{R}}\right)
(U^{R\dagger}G^RU^R)_{e\mu},
\qquad
g_4\simeq \frac{g^{L}}{m_{KK^2}}\left(f-\frac{g^{L}_{1e}}{g^{L}}\right)
(U^{L\dagger}G^LU^L)_{e\mu},
\nonumber \\
g_5\simeq \frac{g^{L}}{m_{KK^2}}\left(f-\frac{g^{L}_{1e}}{g^{L}}\right)
(U^{R\dagger}G^RU^R)_{e\mu},
\qquad
g_6\simeq \frac{g^{R}}{m_{KK^2}}\left(f-\frac{g^{R}_{1e}}{g^{R}}\right)
(U^{L\dagger}G^LU^L)_{e\mu},
\end{eqnarray}
\noindent 
where, in each case,  the first term comes from  $Z$ exchange and the second one is from 
direct KK exchange. The latter can be neglected since 
$f\gg g^{L,R}_{1e}/g^{L,R}$, just as for the case with three 
generations~\cite{Agashe:2006iy}. 

\begin{figure}[h]
\begin{minipage}{\linewidth}
\centering
\psfrag{xAxis}[cb]{(a)~~~$R(\mu^-\rightarrow e^-e^+e^-)$}
\psfrag{yAxis}[cb]{$R(\mu^- -e^-)_{conv.}$}
\includegraphics[width=.475\textwidth,height=2.5in]{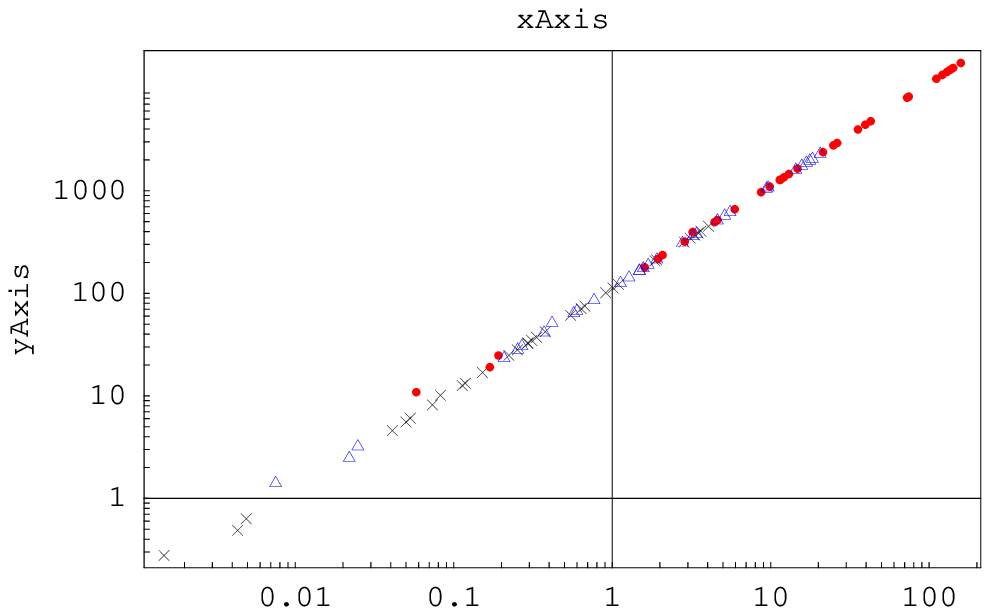}
\quad
\psfrag{xAxis}[cb]{(b)~~~$R(\mu^-\rightarrow e^-e^+e^-)$}
\includegraphics[width=.475\textwidth,height=2.5in]{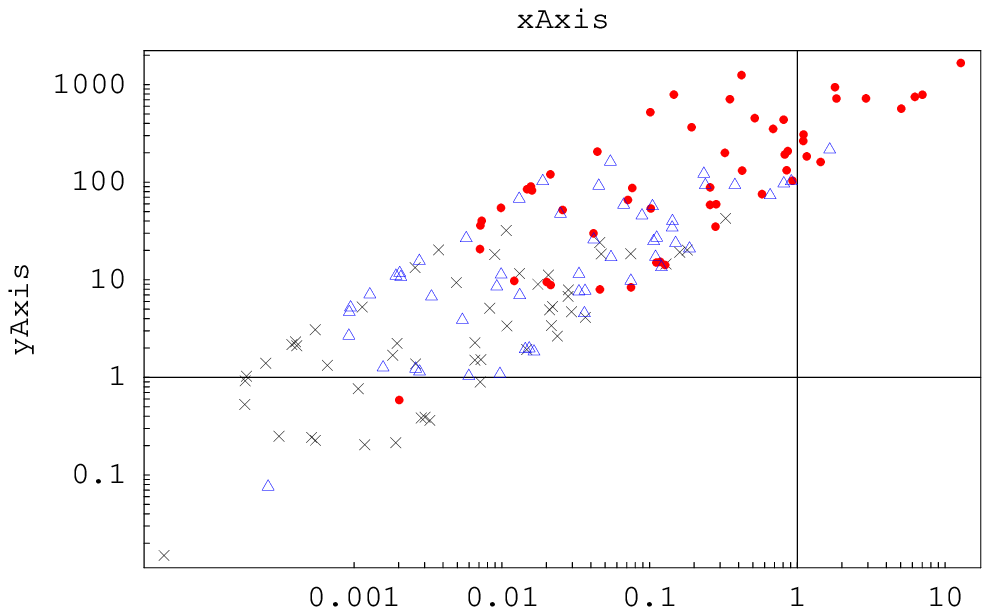}
\end{minipage}
\caption{\em 
Predictions for $\mu^-\to e^-e^+e^-$ 
and $\mu^-e^-$ conversion normalized to  the experimental limits. 
The points correspond to different sets of 5D parameters 
compatible with lepton flavor data and satisfying  the constraints from 
$\mu\to e\gamma$ in (\ref{boundmuegamma}). 
The dots (red), triangles (blue) and crosses (black) correspond to $m_{KK}=2.4,4,6$ TeV. 
The horizontal and vertical lines are the experimental bounds. On the left we 
show the results for the region of the parameter space defined in Section~\ref{sec-neutrinos}. 
On the right we show the results when the localization is 
chosen so that left-handed  and right-handed contributions are of the same order. 
The dispersion in this case is due to different sets of 
Yukawas $\sim{\cal O}(1)$ and to small variations in the 5D fermion masses: $\Delta^i\sim 0.01$.}
\label{Fig-FCNC}
\end{figure}

The  experimental bounds can now be used constrain $m_{KK}$. 
Making use of the naive estimates of Eqs.~(\ref{gZviolatingmuL}-\ref{gZviolating4R}) and the results of~(\ref{matchg3-6}) in Eqs.~(\ref{mu-eee}) 
and~(\ref{mu-e}), we obtain:
\begin{eqnarray}
BR(\mu^-\to e^-e^+e^-): \ m_{KK}\gtrsim 4 \ {\rm TeV} \ ,\qquad
B(\mu^--e^-)_{\rm conv.}: \ m_{KK}\gtrsim 6 \ {\rm TeV} \ .
\end{eqnarray}
A more precise statement about the experimental constraints on the KK scale can be obtained scanning over the parameter space 
defined in Section~\ref{sec-neutrinos}, using the benchmark point of
Eq.~(\ref{becnchmark1}). 
The results are in Figure~\ref{Fig-FCNC}~(a), 
where we show the predictions of the 
model for $\mu^-\to e^+e^-e^-$ and $\mu^--e^-$ conversion in terms of the ratios to the experimental bounds, $R(\mu^-\to e^-e^+e^-)$ and 
$R(\mu^--e^-)_{conv.}$. The points in the plot correspond to different sets of anarchic 5D Yukawas that reproduce 
the observed spectrum and mixings and also satisfy the $\mu^-\to e^-\gamma$ bounds of 
(\ref{boundmuegamma}). We have also allowed a small violation of universal localization 
for the left-handed light leptons, $\Delta^i_{l,\nu}\sim 0.01$.

The $90\%$~C.L. region allowed by both experiments is the lower left corner.  
Note that for $m_{KK}=2.4$ TeV (red points) almost no configuration satisfies 
the experimental constraints, whereas for $m_{KK}=6$ TeV (black crosses) a sizable 
number of the solutions lie within the limits. The most stringent constraint comes 
from $\mu^--e^-$ conversion. If we only consider the bound from  $\mu^-\to e^-e^+e^-$,   
a portion of the parameter space with $m_{KK}=2.4$ TeV is allowed. 

Figure~(\ref{Fig-FCNC})~(a) points to $m_{KK}\gtrsim 6$ TeV,
introducing a little hierarchy that renders the model somewhat unnatural. 
However, the naive choice of parameters resulting in Figure~(\ref{Fig-FCNC})~(a)
is not an optimal one. As noted above,  the largest contributions arise from the left-handed 
flavor violating interactions, at least one order of magnitude 
larger than the right-handed ones.
It is possible to decrease the amount of  flavor violation in the left-handed sector 
by increasing the UV localization of the left-handed light leptons. 
For instance, if we consider $c_l\sim 0.65$, since $f(0.65)/f(0.6)\sim 1/5$, 
this would result in 
a suppression factor of $\sim 1/25$ in the contribution of $e^4_L$ to 
$g^L_{e\mu}$. Also, for larger $c^i_l$, the KK couplings $g^L_{1i}$ become closer 
to the universal value (obtained for $c^i_l=\infty$), and the cancellation 
in $g^L_{1i}-g^L_{1e}$ becomes more efficient. 
Of course, in order to obtain the observed charged lepton masses, we have to increase the 
IR localization of the right-handed light leptons, increasing at the same time the 
size of the flavor violation in the right-handed sector. 
Roughly speaking we increase $g^R_{e\mu}$ by a factor equal to the one 
suppressing $g^L_{e\mu}$. Therefore, we reach the minimum when the flavor 
violation in the left and right-handed sectors are of the same order. 
By playing with the localization in this way, we can decrease (increase) 
$g^L_{e\mu}$ ($g^R_{e\mu}$) by a factor $\sim4$, lowering the contribution to  
$\mu^--e^-$ conversion by one order of magnitude. 
More specifically, we want the size of the left-handed and right-handed 
contributions to be of the same order
\begin{equation}\label{c-min-fcnc}
\frac{f(c^1_l)f(c^2_l)}{f(c^4_l)^2}\sim \frac{f(-c^1_e)f(-c^2_e)}{f(-c^4_e)^2} \ .
\end{equation}
\noindent 
We can satisfy (\ref{c-min-fcnc}) and obtain the correct spectrum by choosing the 5D fermion masses
\begin{eqnarray}
c^1_l\simeq c^2_l \simeq 0.625 \ , \qquad c_e\simeq -0.68 \ , \qquad c_\mu=-0.575 \ ,
\end{eqnarray}
\noindent and the usual localization for the fourth generation. 
For this configuration the contributions from mixings between the first two generations are
as large as the one from the fourth generation. 
We estimate them to be 
\begin{equation}
(U^{L\dagger}G^LU^L)_{e\mu}\sim (U^{R\dagger}G^RU^R)_{e\mu}\sim 10^{-4} \ .
\end{equation}
With these estimates we obtain:
\begin{eqnarray}
BR(\mu^-\to e^-e^+e^-): \ m_{KK}\gtrsim 2 \ {\rm TeV} \ ,\qquad
B_{\rm conv}(\mu^--e^-): \ m_{KK}\gtrsim 4 \ {\rm TeV} \ .
\end{eqnarray}
\noindent
In Fig.~\ref{Fig-FCNC}~(b) we show the results of the scan for this configuration. 
As expected, the constraints now admit solutions with lighter values of $m_{KK}$, 
as low as $m_{KK}\sim 2.4$ TeV.

In addition to optimal localization, there is another way to suppress lepton flavor violation.
This can be achieved with certain choices of embedding of the lepton sector in the
5D gauge theory. 
Although up to now we have not addressed this point, 
we must make such choice in order to define the phenomenology of the model since the 
5D gauge symmetry is 
$SU(2)_L\times SU(2)_R\times U(1)_X$, and not just the SM electroweak gauge sector.
This FCNC-suppressing mechanism is independent of the localization of the zero modes. 
Instead, we will show in the next section that there is a symmetry that protects 
the $Z$ interactions of the left-handed fermions~\cite{Agashe:2006at,Agashe:2009tu}.

\subsection{Custodial symmetry for flavor violating $Z$ couplings}
As promised in the previous section, choosing the embedding of the
lepton sector in the 
5D bulk gauge theory can significantly 
relax the lepton flavor-violation bounds, eliminating the need for a higher KK mass scale. 
By properly choosing the representation of the leptons under the 
5D gauge symmetry, it is possible to protect 
the $Z$ couplings from shifts generated by its mixing with the KK 
resonances~\cite{Agashe:2006at}.
It is generally the case that if the 5D couplings of the SU(2)$_{L,R}$
gauge groups are 
taken to be equal and the fermion involved satisfies $T^{3L}=T^{3R}$, 
there will be  a $P_{LR}$ 
custodial symmetry. 
In the present case, this symmetry can be realized by embedding the 
left-handed doublet in a $({\bf 2},{\bf 2})_0$, protecting 
the left-handed coupling of the charged leptons, such as in  Models 4 and 5 in the previous section. 
Note that we can not protect the $\nu_L$ simultaneously, 
because $T^{3L}=-T^{3R}$ for the $\nu_L$ in this case. 
Instead we would have to choose a $({\bf 2},{\bf 2})_{-1}$ to protect the $\nu_L$ coupling.
To protect the coupling of the right-handed fermions, it is 
possible to invoke also a $P_C$ symmetry, realized when $T^{3L}=T^{3R}=0$. 
Thus we can choose a $({\bf 1},{\bf 1})_{-1}$ or 
$({\bf 1},{\bf 3})_{-1}$ for $e_R$, as in Models 2 and 3, respectively. It is not 
possible to protect both, the left-handed and right-handed couplings of charged 
leptons at the same time, because they have different $X$ charges.

We showed in the previous section that, for the region of the parameter space chosen 
in Section~\ref{sec-neutrinos}, it is necessary to decrease the flavor-violating 
effects for the {\em left-handed charged} leptons. 
This can be done in Embeddings 4 and 5. 
Therefore, the flavor-violating $Z$ couplings 
in the left-handed  sector are suppressed, and we can lower the KK scale 
(see Ref.\cite{Agashe:2009tu} for related discussions). 
Although we have not made a detailed analysis, we expect that this choice of embeddings would alleviate the tension from FCNC, 
allowing for a considerably lower value of $m_{KK}$.
 
Notice that the contributions to FCNC arising from direct KK exchange 
(as opposed to the effects from mixing between $Z^{(0)}$ and $Z^{(n)}$) 
are still present. 
However, these effects are suppressed by a factor $\sim 1/f$, 
and we expect this contribution to be sub-dominant.

\section{Phenomenology}\label{sec-phenomenology}
Many of the lepton properties in this framework are independent of the embeddings 
on the higher dimensional gauge group, and for this reason they are robust predictions if there is a fourth 
generation of leptons in 5D. In the following we will discuss some of these generic properties, 
and then we will comment on some model-dependent properties.

\subsection{Light KK Spectrum}
\label{sec-lightKK}
An important prediction of the model is the presence of light KK leptons of the fourth generation (charged and neutral), 
lighter than the gauge KK modes~\cite{Contino:2006qr}. This can be understood by the following reasons. 
The large 5D gauge symmetry (larger than the SM one) leads to 5D fermions transforming under the full gauge transformations. 
To avoid an excess of zero-modes some of the partners of $e^4$ and $\nu^4$ have $(\pm,\mp)$ boundary conditions. 
For a 5D mass $-1/2\lesssim c\lesssim 1/2$ the mass of the first KK mode of a right-handed fermion with $(-,+)$ boundary conditions can be approximated by:
\begin{equation}
\label{mKK}
m_{\rm{KK}}\simeq A\,k e^{-kL}\,\sqrt{\frac{1}{2}-c} \ ,
\end{equation}
where $A\sim 2$ is approximately constant. Thus, the mass of the first right-handed KK mode is parametrically smaller than the mass of a gauge KK mode, 
with the suppression  given by $\sqrt{1/2-c}$. For a left-handed fermion with $(-,+)$ boundary conditions a similar situation holds changing $c\to -c$.

As discussed in Section~\ref{model}, in order to obtain heavy leptons for the fourth generation, 
the 5D masses have to be in the range: $-1/2\lesssim c^4_{l,\nu,e}\lesssim 1/2$. Moreover, since the see-saw 
mechanism has a strong dependence with $c_\nu$, we need $c^4_\nu\gtrsim0$ to avoid a new light neutrino, experimentally excluded.

From the above arguments and the lepton embeddings of Section~\ref{sec-embedding} we obtain that 
the first KK modes arising from the SU(2)$_R$ partners of $e^4$ and $\nu^4$ are light. This is  a consequence of the large 5D gauge symmetry 
and  the heaviness of the fourth generation\footnote{A similar effect is present in the quark sector for fields 
with zero-modes giving rise to heavy fermions.}. In terms of the 
4D dual description of the theory the light fermionic resonances arise as a consequence of the large global symmetry 
of the strong sector and the fourth generation being almost composite states.

We will consider the different embeddings of the previous section to study the range of masses of the light 
KK fermions. As we see below, the existence of these light states is a generic property present in all embeddings. 
Although we only show the results for Models 1 and 5, that have some important differences, we have explicitly checked that the range of 
masses is similar for all the embeddings of Section~\ref{sec-embedding}. 
We have checked in our calculations that the effects arising form generation mixing are not larger than $25\%$, thus we will neglect the 
mixings between generations for this analysis, and we will consider just the fourth generation that will lead to light KK fermions.
For the purpose of this simplified discussion, we will 
consider that the Higgs is localized in the IR boundary with a $\delta$-function. 

\subsubsection{Model 1} 
After EWSB, there are two towers of leptons: one tower of charged fermions with a light KK mode whose mass is controlled 
by $c^\nu$ (note that $e'_R(-+)$ has the appropriate boundary conditions), and one tower of neutral fermions with a 
light KK mode whose mass is controlled by $c^e$ ($\nu^{''}_R(-+)$ has the appropriate boundary conditions).
The spectrum of charged fermions is given by
\begin{eqnarray}
{\rm zeroes}\left[\frac{f^R_{\alpha_l}}{f^L_{\alpha_l}}+({\cal M}^e z_1)^2
\left(\frac{f^R_{\alpha_e}}{f^L_{\alpha_e}}+\frac{f^R_{\alpha_\nu}}{f^L_{\alpha_\nu}}\right)\right]|_{z_0} \ ,
\end{eqnarray}
where $f^{L,R}_{\alpha_\psi}$ is the KK wave function of the left (right)-handed charged fermion $\psi$ and ${\cal M}^e$ is the Dirac mass from a 
localized IR Higgs. The spectrum of neutral fermions is given by a similar equation. 

\begin{figure}[h]
\centering
\psfrag{xAxis}{$c_e, c_{\nu}$}
\psfrag{yAxis}{$m_{\nu} (c_e), m_{e} (c_{\nu})$ [TeV]}
\includegraphics[width=.6\textwidth]{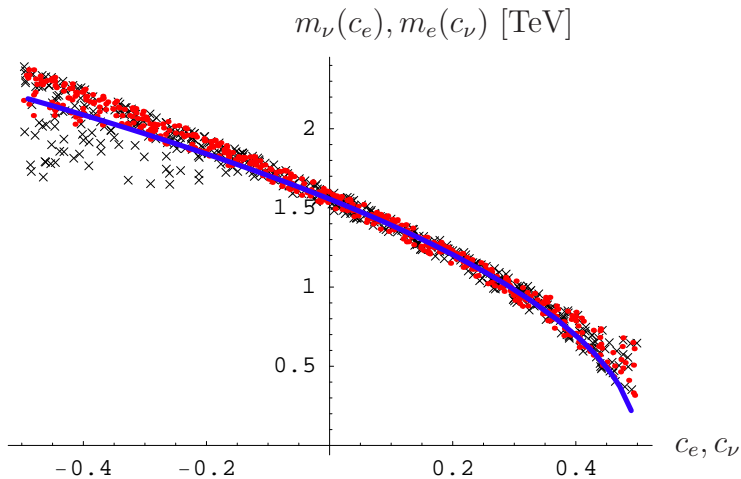}
\caption{\em Model~1. Red dots (black crosses)   correspond to the mass of the lightest 
charged (neutral) KK fermion as a function of $c_\nu$ ($c_e$). 
The blue line is the prediction given by Eq.~(\ref{mKK}). We have made a random scan with $-0.5\leq c^4_{l,e\nu}\leq 0.5$ 
and we have fixed ${\cal M}^{e,\nu}=0.25\,k\,e^{-kL}$.}
\label{Fig-mass-me1}
\end{figure}

\begin{figure}[h]
\centering
\psfrag{xAxis}{$m_{e^{(1)}}$ (TeV)}
\psfrag{yAxis}{$m_{\nu^4}$ (TeV)}
\includegraphics[width=.6\textwidth]{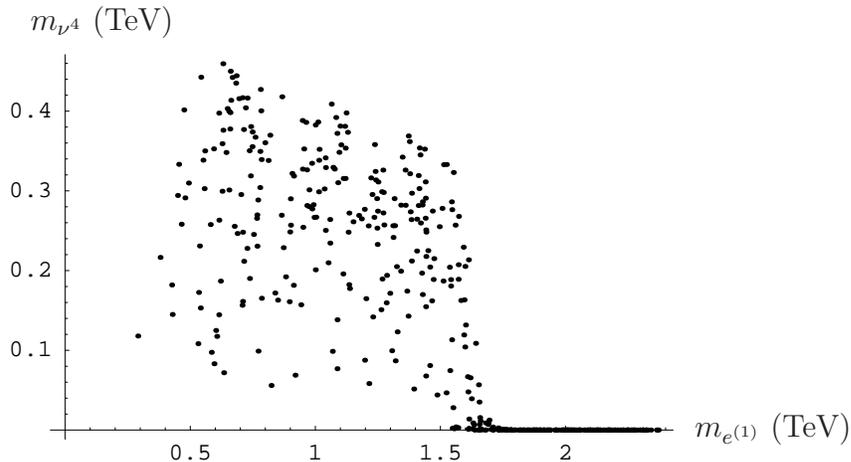}
\caption{\em Model~1. Mass of the zero mode $\nu^4$ as a function of the mass of the first KK 
charged fermion $e^{(1)}$. We have varied $-0.5\leq c^4_{l,e,\nu}\leq 0.5$ and we have fixed ${\cal M}^{e,\nu}=0.25/z_1$.}
\label{Fig-mn0-me1}
\end{figure}

Figure~\ref{Fig-mass-me1} shows the mass of the lightest charged (neutral) KK mode as a function of $c^\nu$ ($c^e$), 
and the prediction given by Eq.~(\ref{mKK}) with $A=2.3$. We have made a random scan over the 5D masses constraining 
$-0.5\leq c^4_{l,e\nu}\leq 0.5$ and we have fixed ${\cal M}^{e,\nu}=0.25\,k\,e^{-kL}$, that corresponds to Yukawas of order one.
In Figure~\ref{Fig-mn0-me1} we show the mass of the lightest charged KK versus the mass 
of the zero mode $\nu^4$ using the same scan as in the previous plot. 
We can clearly see that, since a heavy $\nu^4$ requires $c^\nu_4\gtrsim 0$, 
this model predicts a charged KK not heavier than $\sim1.5\,k\,e^{-kL}$ ($\simeq1.5$~TeV 
for $k\,e^{-kL}\simeq1$~TeV).

The new fermions give contributions to the oblique parameters. The most important constraint comes from the contributions of the 
would be zero modes of the fourth generation to the $T$ parameter, that is constrained by 
electroweak precision measurements to be not much larger than $T\sim0.1$\footnote{The precise value of the bound on $T$ depends on the value of the other 
electroweak observables in the model.}. Fixing the leptonic contribution to be $\Delta T= 0.1$ we constrain the isospin splitting to 
be $\Delta m=|m_{\nu^4}-m_{e^4}|\sim 74$ GeV. This result has important consequences for the phenomenology, since $\Delta m<m_W$ 
and then the leptons of the fourth generation will preferentially decay to $W+l$, with $l$ labeling the SM leptons. However this 
result depends on the precise value of $\Delta T$. For instance, for $\Delta T\sim 0.3$ we obtain $\Delta m\sim 130$ GeV$>m_W$. Thus, electroweak precision constraints 
prefer a small isospin splitting, although we can not completely exclude the possibility $\Delta m>m_W$. The constraints on the $T$ 
parameter exclude a region of the parameter space of the 5D model. Demanding $\Delta T=0.1$ we keep $\sim 60\%$ of the parameter 
space in Model~1, where we have considered only the region of the parameter space where $\nu^4$ and $e^4$ are heavy enough. Thus, the result 
has a mild dependence on this lower limit. For example, for $m_{\nu^4,e^4}>50$ GeV (250 GeV) we keep $51\%$ ($72\%$) of the parameter space. 

A similar analysis can be done for Models 2-4. In all these models we need $c_\nu^4\gtrsim 0$ to obtain a heavy $\nu^4$, therefore light KK modes will arise from the SU(2)$_R$ partners of $\nu_R$, that is embedded in a triplet of SU(2)$_R$. In Models 2 and 3 there is a light mode with the electron quantum numbers, $Q=-1$, and there is also an exotic state $\chi$ with $Q=-2$, whereas in Model~4 $\chi$ has $Q=+1$. 
Since the left-handed doublet is embedded in a bi-doublet of SU(2)$_L\times$SU(2)$_R$, there are also light charged and exotic KK modes associated to the SU(2)$_R$ partners of $L_L$, {\it i.e.:} $L'_L$. In this case their masses are controlled by $c_l^4$. In Models 3 and 4 the right-handed electron is embedded in a triplet of SU(2)$_R$, thus there are also light states associated to the SU(2)$_R$ partners of $e^4_R$, with the spectrum controlled by $c_e^4$. 
After the Higgs acquires a VEV, the fields with the same electric charge are mixed, thus the mass of the light states can depend on the parameters $c_{l,e,\nu}^4$, according to Eq.~(\ref{mKK}). Let us discuss briefly Model 2, as an example: there are light neutral sates for $c^l$ near $-1/2$, and there are light charged and exotic states for $c^l$ or $-c^\nu$ near $-1/2$ (the last condition is similar to Model 1). As discussed previously, to obtain a heavy fourth generation we need $-1/2\leq c_{l,e}^4\leq 1/2$ and $0\lesssim c_\nu^4\leq 1/2$, thus the last condition guarantees the existence of light charged and exotic states associated to the $\nu_R$ partners, and the other conditions can give an extra suppression in the mass of the first KK.

\subsubsection{Model 5} 
\begin{figure}[h]
\begin{minipage}{\linewidth}
\psfrag{xAxis}{$m_{e_4}$(TeV)}
\psfrag{yAxis}{$m_{KK}$(TeV)}
\epsfig{file=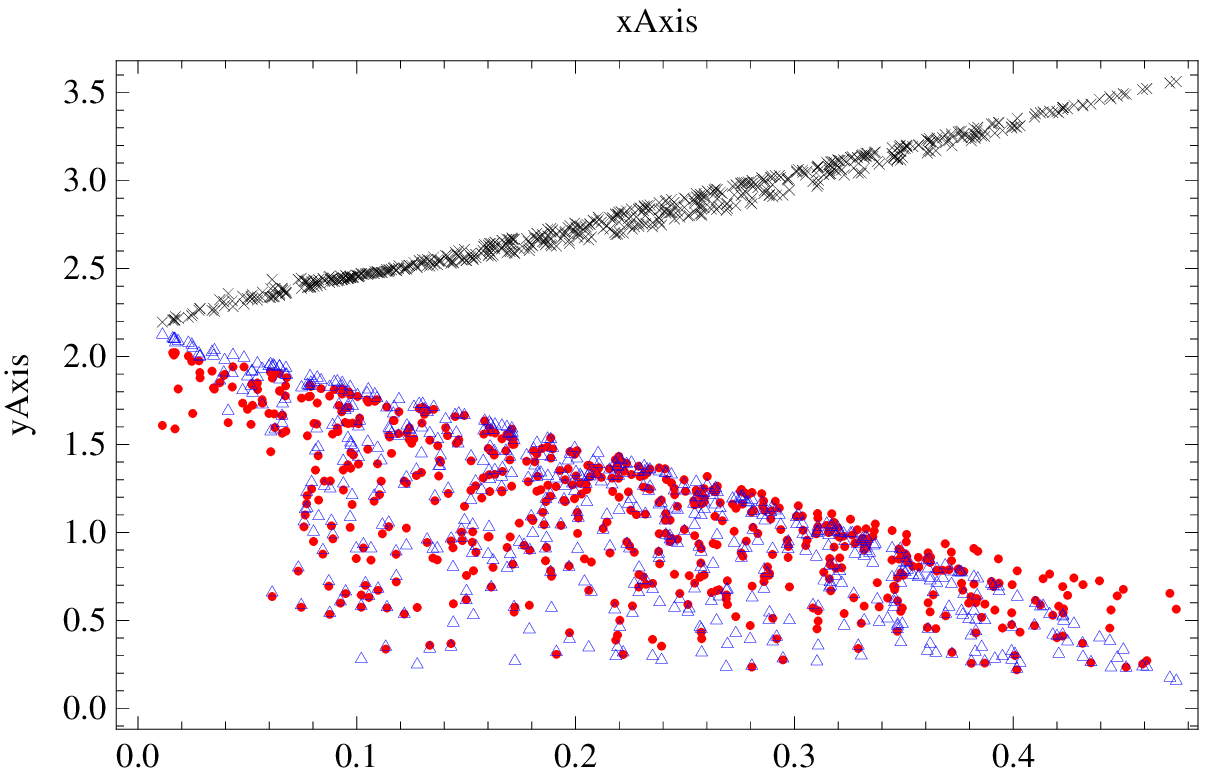,width=0.498\linewidth}
\quad
\psfrag{xAxis}{$m_{\nu_4}$(TeV)}
\epsfig{file=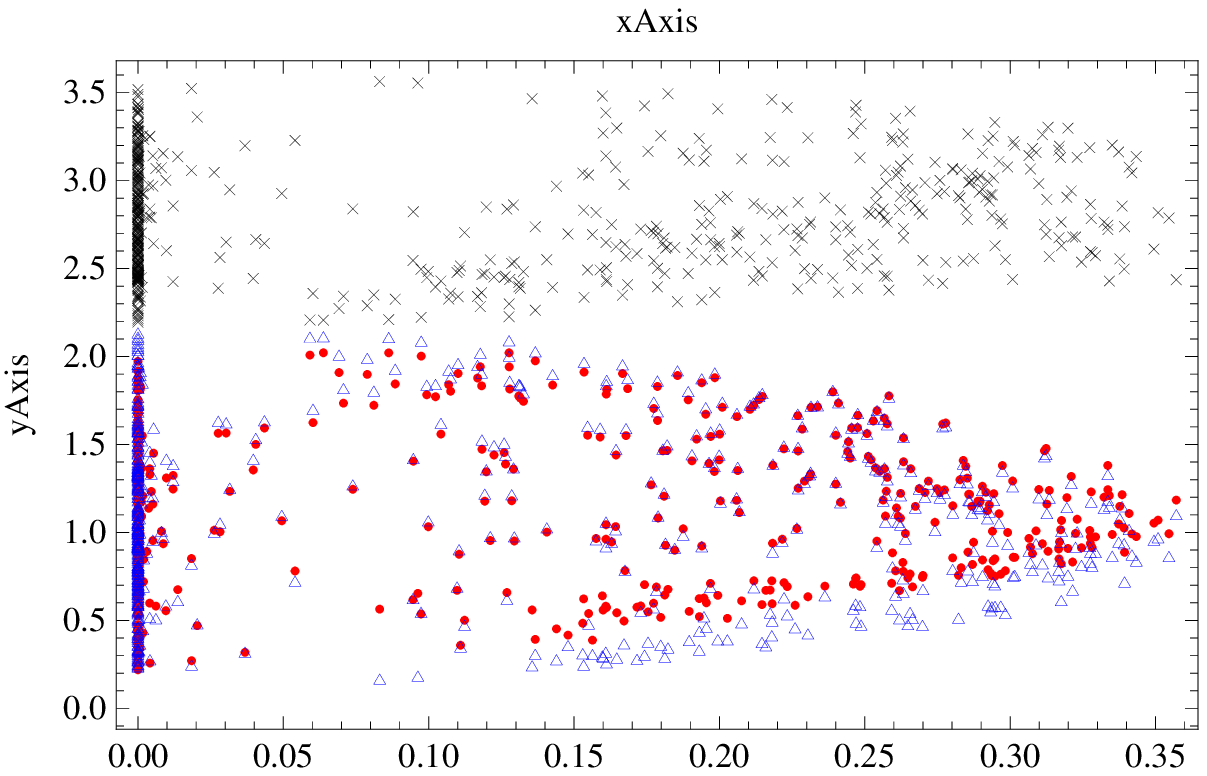,width=0.498\linewidth}  
\end{minipage}
\caption{\em Model~5. Crosses (black), dots (red) and triangles (blue) correspond respectively to the mass of the lightest charged, neutral, exotic KK fermions. 
On the left (right) we show the KK mass as a function of the mass of the zero-mode 
electron (neutrino) of the fourth generation $e^4$ ($\nu^4$). 
We have made a random scan over the parameter space with $-0.5\leq c^4_{l,e,\nu}\leq0.5$.}
\label{Fig-model5}
\end{figure}

We consider this model separately, due to its somewhat distinct features. 
In this case, $\xi^\nu$ is a singlet, implying that fermions with $(-,+)$ boundary conditions 
arise from $\xi^l$ and $\xi^e$. As previously, to obtain a heavy $e^4$ we need $-1/2\lesssim c^4_{1,e}\lesssim 1/2$. 
Thus there is a region of the parameter space leading to light
fermionic KK resonances ($-1/2\lesssim c^4_l\lesssim 0$ 
and/or $0\lesssim c^4_e\lesssim1/2$), and another region 
$0\lesssim c^4_l\lesssim1/2$ and $-1/2\lesssim c^4_e\lesssim0$ 
where we can still obtain a heavy fourth generation and the suppression in the KK mass is not so large. 
Thus, once we select the region of the parameter space leading to
heavy leptons,  Model 5 will give, in general, 
light fermionic resonances, but there are some regions where the
suppression is small 
(roughly $20\%$ of the allowed region). In Figure~\ref{Fig-model5} we plot the KK mass in terms of the masses
of $e^4$ and $\nu^4$, using ${\cal M}_{IR}=0.25 k $ for the
IR-localized Majorana mass.
Besides the usual charged and neutral states there is another state with $Q=+1$. 
After EWSB the mass eigenstates arise from  mixing between the 
upper component of $L'$ and $\chi$ in Eqs.~(\ref{xil-model4}) and~(\ref{xinu-model5}).

Note that in this model there are no fields with $(-+)$ boundary conditions and $Q=-1$, 
thus there is no light KK associated to the electron, as can be seen in Figure~\ref{Fig-model5}.
The contributions of the zero modes to the $T$ parameter are similar to Model 1.

\subsection{Couplings of Fourth Generation Leptons}
\label{subsec:couplings}
In this section we study some features of the interactions involving  fourth-generation leptons  
that have an impact on their collider phenomenology. 
Since the IR localization of the fourth generation is larger than the SM one, the following relation is satisfied: 
$M^{\nu,44}_{LR}\gg M^{\nu,i4}_{LR}$ for $i=1,2,3$, see Eq.~(\ref{v45}). 
Using this result, to leading order we can write $U^\nu$ as:
\begin{equation}\label{Unu-aprox}
U^\nu\simeq\left[\begin{array}{cc}
U^\nu_{3\times3}&0_{3\times2}\\
0_{2\times3}&\left[\begin{array}{cc}1/\sqrt{2}&1/\sqrt{2}\\1/\sqrt{2}&-1/\sqrt{2}\end{array}\right]
\end{array}\right] \ .
\end{equation}
Inserting Eq.~(\ref{Unu-aprox}) in (\ref{LW}) we obtain, at this level of approximation:
\begin{equation}\label{LW1}
{\cal L}_{CC}\supset\sum_{\alpha=1,2,3}\frac{g}{\sqrt{2}}W_\mu^{+}\bar\nu^4_L \gamma^\mu U^{L}_{4\alpha}e^\alpha_L+ {\rm h.c.} \ .
\end{equation}
\noindent 
Therefore the $\nu^4$ decay is mostly determined by $U^{L}_{4a}$, $a=1,2,3$, 
i.e. only the mixings in the charged sector are important. Moreover, since $c^i_l\simeq c_l$ for $i=1,2,3$, the model predicts that all the branching ratios in the 
decay $\nu^4\to lW$ are approximately equal. 
We have verified this result in our numerical scan over the parameter space. This has important consequences for collider searches.

An important aspect of these models is the strong 
coupling between the fourth generation zero modes and the KK excitations of the gauge bosons. 
Thus, the zero-mode leptons of the fourth generation are expected to be strongly coupled to the KK excitations of the electroweak gauge bosons. The couplings of the zero-mode leptons to the 
KK excitations of the SM gauge bosons: $W^{(1)\pm}, Z^{(1)}$ and $\gamma^{(1)}$ are the same as the SM gauge couplings of the lighter three generations, up to the enhancement resulting from the zero-mode IR localization, and small flavor-violating corrections. 
On the other hand, the model has one more neutral gauge boson, the $Z^{'(1)}$, which has no zero mode since it corresponds to the 
broken-generator combination in the breaking $SU(2)_R\times U(1)_X\to U(1)_Y$. The couplings of zero-mode leptons, again in units of 
the localization enhancement factor, are given in Table~\ref{tab:couplings}, where $g_R$ and $g_X$ are the $SU(2)_R$ and $U(1)_X$ 
5D gauge couplings divided by $\sqrt{L}$ so as to render them dimensionless.
\begin{table}[h]
\begin{center}
\begin{tabular}{|c|c|c|c|c|c|}
\hline
Zero Mode & E1 & E2 & E3 & E4 & E5  \\
\hline
~     &               &                                 &                                             & &\\
$N_L$ & $\frac{1}{2}\,\frac{g_X^2}{\sqrt{g_R^2+g_X^2}}$ &  $\frac{g_R^2/2+g_X^2}{\sqrt{g_R^2+g_X^2}}$ & $\frac{g_R^2/2+g_X^2}
{\sqrt{g_R^2+g_X^2}}$ &
$-\frac{1}{2}\,\frac{g_R^2}{\sqrt{g_R^2+g_X^2}}$ & $-\frac{1}{2}\,\frac{g_R^2}{\sqrt{g_R^2+g_X^2}}$ \\
~     &               &                                 &                                             & &\\
\hline
~     &               &                                 &                                             & &\\
$N_R$ & $\frac{1}{2}\,\sqrt{g_R^2+g_X^2}$ & $\sqrt{g_R^2+g_X^2}$ & $\sqrt{g_R^2+g_X^2}$ & 0 & 0 \\
~     &               &                                 &                                             & &\\
\hline
~     &               &                                 &                                             & &\\
$E_L$ &  $\frac{1}{2}\,\frac{g_X^2}{\sqrt{g_R^2+g_X^2}}$ &  $\frac{g_R^2/2+g_X^2}{\sqrt{g_R^2+g_X^2}}$ & $\frac{g_R^2/2+g_X^2}
{\sqrt{g_R^2+g_X^2}}$ &
$-\frac{1}{2}\,\frac{g_R^2}{\sqrt{g_R^2+g_X^2}}$ & $-\frac{1}{2}\,\frac{g_R^2}{\sqrt{g_R^2+g_X^2}}$ \\
~     &               &                                 &                                             & &\\
\hline
~     &               &                                 &                                             & &\\
$E_R$ & $\frac{1}{2}\,\frac{g_X^2 - g_R^2}{\sqrt{g_X^2+g_R^2}} $ & $\frac{g_X^2}{\sqrt{g_X^2+g_R^2}}$ & $\frac{g_X^2}{\sqrt{g_X^2+g_R^2}}$ 
& $-\frac{g_R^2}{\sqrt{g_X^2+g_R^2}}$ & $-\frac{g_R^2}{\sqrt{g_X^2+g_R^2}}$ \\
~     &               &                                 &                                             & &\\
\hline
\end{tabular}
\end{center}
\caption{\em Couplings of zero-mode leptons to the Z', in units of the dimensionless 
enhancement from localization.}
\label{tab:couplings}
\end{table}

In addition, there are two charged $SU(2)_R$ KK gauge bosons, $R^{\pm (1)}_\mu$, resulting 
in  couplings of the zero-mode leptons to KK leptons. 
Of particular interest are the charged couplings of zero-mode fourth-generation leptons to a light KK lepton. For instance, in Model~1 the zero-mode 
right-handed neutrino $\nu_R(++)$ in (\ref{emb1}) couples through the charged right KK mode $R_\mu^\pm$ to the light KK mode $e'_R(-+)$ with strength $g_R/\sqrt{2}$.
This is also the case for the coupling between the  zero-mode right-handed electron $e_R(++)$, which couples through the charged right current to its $SU(2)_R$ 
partner $\nu^{''}_R$. 
Similar charged couplings exist in Models~2, 3, 4 and 5, whenever the zero-mode right-handed neutrino $\nu_R(++)$ or electron $e_R(++)$ belong to a triplet of
$SU(2)_R$. These interactions are of interest since, as shown in Section~\ref{sec-lightKK}, the right-handed KK leptons with $(-+)$ boundary conditions are lighter than 
the typical KK mass scale. Thus, due to the strong coupling of the fourth-generation lepton zero-modes to both the KK gauge bosons and fermions, the 
single production of a lepton KK mode in association with a fourth-generation zero-mode lepton is a potentially important signal for these models.

Given the couplings above, we can study the phenomenology of fourth-generation leptons at the LHC in the models presented here. 
The fourth-generation zero-mode leptons will be produced via the s-channel exchange of the SM gauge bosons, as well as the 
KK gauge bosons including the photon, $W$ and $Z$ KK excitations 
$A^{(1)}$, $W^{\pm(1)}$ and $Z^{(1)}$, as well as the KK excitation coming from the 
combination of the generators $T^3_R$ and $X$ broken by the boundary conditions on 
the IR brane, $Z^{'(1)}$. They will also be produced through 
an s-channel Higgs. 

Finally, as  it was shown in Section~\ref{fcnc}, 
there are lepton flavor violating interactions with the $Z$ and the heavier neutral resonances. 
These interactions allow, in principle,  the production of a charged lepton of the fourth generation 
$e^4$ together with a light SM charged lepton through an s-channel neutral vector $Z$ 
or $Z^{(n)}$. Since all the left-handed light leptons have equal localization, 
the interaction $\bar e^4_L Z e^i_L$, $i=1,2,3$, has the same strength for any flavor 
(a similar result holds for $Z^{(n)}$). 
Thus the production process $Z,Z^{(n)}\to \bar e^4_L e^i_L$ and the 
decay $e^4_L\to Ze^i_L$ are almost flavor independent. On the other hand, 
the right-handed interactions $\bar e^4_R Z e^i_R$ are stronger for heavier 
leptons $e^i$, thus for neutral right-handed currents the dominant production 
mechanism is $Z,Z^{(n)}\to \bar e^4_R \tau_R$ and the dominant neutral decay is 
$e^4_R\to Z\tau_R$. The flavor-violating couplings, however, are likely to be too small to be observed at the early 
stages of the LHC~\cite{DePree:2009ed}. 

\subsection{$\nu^4_R$ phenomenology}
Although a detailed phenomenological 
study will be done elsewhere~\cite{lepton2}, we make here some general 
remarks regarding the right-handed zero-mode neutrino, $\nu^4_R$.
We start with a brief discussion of the production of $\nu^4_R$ at hadron colliders.
In Models 1, 2 and 3 right-handed neutrinos can be pair-produced through 
their couplings to $Z'^{(1)}$, shown in Table~\ref{tab:couplings}.
This is not the case in Models 4 and 5. However, 
since the Yukawa couplings of the fourth-generation neutrinos are
large, the $\nu^4_R$ can be produced as a decay product of the Higgs, through
$g g \to h \to \nu^4_L \nu^4_R $.
For a heavy Higgs such that $m_h>2m_{\nu^4}$, with
$m_{\nu^4}\sim{\cal O}(300)$ GeV, we obtain a branching ratio 
$BR(h\to\nu^4_L \nu^4_R)\sim (2-5)\%$, for $m_h\sim (650-900)$~GeV. 
Adding the fact that 
the presence of a full fourth generation increases significantly
the Higgs production cross
section by gluon fusion over the usual SM value~\cite{fourgen}, this production 
mechanism for $\nu^4_R$ cannot be neglected, 
not just  in Models 4 and 5, but also in Models 1-3~\cite{lepton2}.  

Finally, in all cases $\nu^4_R$ decays promptly through the
Dirac-mass mixing with $\nu^4_L$ into a $W$ and a light charged
lepton. Additionally, 
in embeddings  1-4  (see Section~\ref{sec-embedding}),
where the $\nu_R$ is charged under the SU(2)$_R$, the $\nu^4_R$ decay
can proceed through a virtual charged KK vector $R^{\pm(n)}$ and a
virtual KK charged fermion $e^{'(n)}$ of the fourth generation.
$R^{\pm(n)}$ mixes via Higgs mass insertions with the zero mode
$W^{(0)}$ and $e^{'(n)}$ mixes with the charged zero-mode fermions
$e^{i(0)}$. However, this modes will be suppressed by KK masses
compared to the one mentioned above.  

\section{Conclusions}
\label{conclusions}
Extensions of the standard model with a fourth generation must address the 
question of why the fourth generation neutrinos are so much heavier than the 
three-generation light neutrinos. In addition, they must satisfy 
constraints from lepton flavor violation and reproduce the pattern of mixings and mass differences of light neutrinos.    
We have addressed these questions in 
a class of models in the context 
of theories with warped extra dimensions.
In this scenario both the charged lepton and the neutrino are heavy as a 
consequence of their localization in the extra dimension. 
They are both IR-localized which results in a large overlap with the Higgs, 
independently of whether this is an elementary scalar or 
an effective degree of freedom resulting from the condensation of fourth-generation quarks. 
The models naturally have a UV-localized Majorana mass term 
which results in a see-saw mechanism only efficient with neutrinos with zero-mode 
wave-functions localized close to the UV brane. 
Thus, a consequence of this setup is that the light neutrinos of the first three 
generations are mostly Majorana particles since they have a significant 
overlap with the UV brane and therefore feel the effect of the Majorana mass term. On the other hand, fourth-generation neutrinos are IR-localized and 
will be mostly Dirac particles, since they have little overlap with the Majorana mass term.

The exception to this is the case where right-handed neutrinos come
from singlets under the bulk gauge symmetries (embedding 5 
in Section~\ref{sec-embedding}). As a consequence, in addition to the
UV-localized Majorana term present in embeddings 1-4, there is an
IR-localized Majorana mass that only affects significantly the fourth
generation neutrinos. The two Majorana components are split with an 
order-one splitting, resulting in a potentially light Majorana
fourth-generation neutrino that could be as light as $100$~GeV.

For all the possible embeddings, we have shown that in order to satisfy the existing 
constraints from neutrino mixings and in combination with the lepton
spectrum the localization of the lepton left-handed zero-mode doublets 
must be almost degenerate for the first three generations, i.e. $c^i_\ell=c_\ell$, for $i=1,2,3$. Deviations from the equality cannot significantly exceed 
$1\%$ in order to avoid a hierarchical mixing pattern in the light neutrino sector. 
Similarly, the localization of the right-handed zero-mode neutrinos of the light three generations is chosen to be the same in order to 
match the see-saw picture with a single suppression scale.  
This suggests the presence of a flavor symmetry acting on UV-localized left-handed leptons and right-handed neutrinos. 
We do not need to impose that the bulk masses of the right-handed charged leptons be the same for the light generations. This freedom allows us to 
obtain the correct spectrum. 

The presence of flavor-violating interactions leads to several lepton flavor-violating processes. We studied the impact of 
the experimental bounds on lepton flavor violating processes on the parameter space of the model.  
We have found that, for generic embeddings, the KK mass scale is bound to be $m_{KK}\gae 4$~TeV by the experimental limits on 
$\mu^-\to e^-e^+e^-$, whereas considering the rate of $\mu^-\to e^-$ conversion results in $m_{KK}\gae 6$~TeV, 
although there are some small regions of parameter space where the FCNC effects are minimized and $m_{KK}\sim 2.4$ TeV is allowed. 
We have shown that in two of the five embeddings proposed in Section~\ref{sec-embedding}, those we called Model~4 and~5, it is possible to 
evade the lepton flavor violation bounds even with KK masses as low as $m_{KK}\simeq 2.4$~TeV, for larger regions of the parameter space. 
These embeddings realize a custodial symmetry that
protects the $Z$ couplings to leptons. 

The parameter space of the models is almost entirely determined by the bulk mass parameters of the zero-mode leptons. This fixes the 
zero-mode mass spectrum of the fourth generation as well as its couplings to the KK gauge bosons and the Higgs, through the enhancement factor 
resulting from IR localization. This, together with the gauge couplings from Table~\ref{tab:couplings}, can be used to study the phenomenology 
of the fourth-generation lepton sector at colliders. As shown in the Table, these  gauge couplings depend on the embedding of the leptons in the 
5D gauge theory. We considered five different possible embedding in Section~\ref{sec-embedding}.
Although the gauge couplings of the zero-mode leptons to the SM electroweak gauge bosons and their KK excitations do not depend on the 5D embedding, 
their couplings to the $Z^{'(1)}$, the KK mode of the broken generator in $SU(2)_R\times U(1)_X\to U(1)_Y$, do as it can be seen in 
Table~\ref{tab:couplings}. 
Furthermore, in Model~5, the right-handed neutrino has no 5D gauge couplings. Thus, in this embedding, the 
fourth-generation $\nu^4_R$ couples only through the Higgs
sector. This implies that both its production and decay must involve
its Yukawa couplings. The dominant production of $\nu^4_R$ is then
through the s-channel Higgs, with the decay proceeding through the
mass mixing with $\nu^4_L$.

Another important prediction is the appearance of light KK leptons, as described in Section~\ref{sec-lightKK}. As shown in Figures~\ref{Fig-mass-me1}, 
\ref{Fig-mn0-me1} and \ref{Fig-model5}, KK leptons with $(-+)$ boundary conditions can be as light as $0.5~$TeV and are generically considerably lighter than 
the KK gauge bosons. One important consequence is that it is more kinematically favorable to singly produce a KK lepton in association with a zero-mode lepton, 
particularly if this is a fourth generation lepton since the corresponding effective gauge coupling would be enhanced due to its IR localization. 
This and other aspects of the  phenomenology of the lepton sector of a fourth-generation at the LHC, including its production and decay, 
will be studied in detail in a future publication~\cite{lepton2}.

\newpage
\appendix
\section{Perturbative diagonalization of the neutrino mass}
\label{Ap-diagonalization}
$\nu$ contains four generations of neutrinos, we split them in the following way:
\begin{equation}
\nu^t=(\nu^e,\nu^\mu,\nu^\tau,\nu^4)=(\nu^a,\nu^4)\ , \qquad a=1,2,3 \ ,
\end{equation}
\noindent thus Latin indexes label the light generations. Using this notation we can write the Dirac and Majorana mass matrices as:
\begin{equation}
M^\nu_{RR}=\left[\begin{array}{cc} M^{ab}_{RR} & M^{a4}_{RR} \\ M^{4b}_{RR} & M^{44}_{RR} \end{array}\right] \ ,
\qquad
M^\nu_{LR}=\left[\begin{array}{cc} M^{ab}_{LR} & M^{a4}_{LR} \\ M^{4b}_{LR} & M^{44}_{LR} \end{array}\right] \ .
\end{equation}
We integrate out the heavy neutrinos $\nu^a_R$ at tree level, and obtain the effective mass term of Eq.~(\ref{Meffnu}), with $M_{\rm eff}$ given by:
\begin{equation}
M_{\rm eff}=\left[\begin{array}{ccc} 
-M^{ab}_{LR}(M^{bc}_{RR})^{-1}M^{cd}_{RL} & -M^{ab}_{LR}(M^{bc}_{RR})^{-1}M^{c4}_{RL} & M^{a4}_{LR}-M^{ab}_{LR}(M^{bc}_{RR})^{-1}M^{c4}_{RR} \\ 
-M^{4b}_{LR}(M^{bc}_{RR})^{-1}M^{cd}_{RL} & -M^{4b}_{LR}(M^{bc}_{RR})^{-1}M^{c4}_{RL} & M^{44}_{LR}-M^{4b}_{LR}(M^{bc}_{RR})^{-1}M^{c4}_{RR} \\ 
M^{4d}_{RL}-M^{4b}_{RR}(M^{bc}_{RR})^{-1}M^{cd}_{RL} & M^{44}_{RL}-M^{4b}_{RR}(M^{bc}_{RR})^{-1}M^{c4}_{RL} & M^{44}_{RR}-M^{4b}_{RR}(M^{bc}_{RR})^{-1}M^{c4}_{RR}
\end{array}\right] \ .
\end{equation}
We can see that most of the entries of $M_{\rm eff}$ are suppressed by the see-saw mechanism. Moreover, since $\nu^4_R$ is localized towards the IR and the 
right-handed Majorana mass is localized in the UV, $M^{44}_{RR}$ is very suppressed also (one can check that for the range of parameters interesting for 
the phenomenology, $M^{44}_{RR}$ is several orders of magnitude smaller than the entries not suppressed by the see-saw). Therefore, at leading order, 
we obtain an effective theory with two heavy Majorana states described by Eqs.~(\ref{m45}) and~(\ref{v45}). At this level of approximation, there is a 
degenerate subspace o dimension 3 with eigenvalues equal to zero, corresponding to three massless neutrinos. These massless eigenstates have no projection 
on the fifth component at this order. Introducing a spurious infinitesimal $\epsilon$ we can compute the light eigenvalues at leading order:
\begin{equation}
{\rm det}(M^\epsilon_{\rm eff}-\epsilon \lambda)=0 \ , \qquad 
M^\epsilon_{\rm eff}=\left[\begin{array}{cc} \epsilon S^{ij} & M^{i4} \\ M^{4i} & \epsilon T \end{array}\right] \ , 
\qquad i,j=1,\dots 4\ ,
\end{equation}
where we explicitly show the suppressed entries of $M_{\rm eff}$. At leading order in $\epsilon$ ({\it i.e.}: ${\cal O}(\epsilon^3)$) this equation has three solutions. 
Once we obtain the light eigenvalues $\lambda^a$, we can write, at leading order, a explicit expression for the light eigenvectors in terms of the $4\times4$ 
matrix $S$ and the dimension 4 vector $M^{i4}$. For simplicity we split the light eigenvectors isolating the fifth component:
\begin{equation}
v^{a}=\left[\begin{array}{c}v^{a}_i\\v^a_5\end{array}\right] \ , \qquad a=1,2,3 \ , \ i=1,\dots 4 \ ,
\end{equation}
\noindent where $a$ labels the light eigenvectors and $i$ labels the first four components of each vector. Thus $v^a_i$ and $v^a_5$ are given by the 
following 
equations in a space of dimension 4:
\begin{eqnarray}
|v^a_5|^2&=&\left(M^\dagger[(S-\lambda^a)^{-1}]^\dagger(S-\lambda^a)^{-1}M\right)^{-1} \ , \\
v^a_i&=&-v^a_5 [(S-\lambda^a)^{-1}M]_i \ ,
\end{eqnarray}
\noindent where, although we have not explicitly shown the indexes, $M=M^{i4}$ is a vector of dimension 4, and $(S-\lambda^a)^{-1}$ is the inverse of matrix $(S-\lambda^a)$, acting on $M$.

\section{Scan}\label{Ap-numbers}
There is a large set of solutions giving the right spectrum and mixings in the leptonic sector, together with a heavy fourth generation. As discussed in 
Section~\ref{sec-neutrinos}, 
there are also some additional constraints arising from the phenomenology, like $\mu\to e\gamma$. We show here a set of 5D parameters satisfying all these constraints, 
with light neutrino mass splittings given by:
\begin{equation}
\Delta m^2_{\rm sol}\simeq 1.1\ 10^{-4}{\rm eV}^2 \ , \qquad \Delta m^2_{\rm atm}\simeq 2.6\ 10^{-3}{\rm eV}^2 \ .
\end{equation}
We have scanned over the parameter space, considering real symmetric Yukawas and the identity for $\lambda_R$ 
in order to simplify the computation.
The 5D Yukawa matrices are approximately given by:
\begin{eqnarray}
\lambda_e=\left[\begin{array}{cccc}
2.5 & 1.1 & 0.6 & -0.6 \\ 1.1 & 1.1 & -0.5 & -0.6 \\ 0.6 & -0.5 & -0.1 & -1.4 \\ -0.6 & -0.6 & 1.4 & 1.1 \end{array}\right] \ , \ \
\lambda_\nu=\left[\begin{array}{cccc}
-2.4 & 0.8 & -1.1 & -0.6 \\ 0.8 & 3.9 & -1.5 & -1.4 \\ -1.1 & -1.5 & 2.8 & 0.7 \\ -0.6 & -1.4 & 0.7 & 1.1 
\end{array}\right] \ , \ \
\lambda_R= 1_{4\times 4} \ .
\end{eqnarray}

The spectrum of zero modes reproduces the charged leptons of the SM, with heavy leptons: $m_\nu^4\simeq m_e^4=300$~GeV, and 
light neutrinos with masses approximately given by~\cite{Csaki:2008qq}:
\begin{equation}
m_\nu^1=18 \ {\rm meV} \ , \qquad m_\nu^2=53 \ {\rm meV} \ , \qquad m_\nu^3=54 \ {\rm meV} \ . 
\end{equation}
The corresponding mixing matrix $V$ is given by:
\begin{eqnarray}
V=\left[\begin{array}{cccc}
0.82 & -0.34 & 0.46 & -0.04 \\ 0.57 & 0.39 & -0.72 & 0.03 \\ 0.06 & 0.85 & 0.52 & 0.01 \\ -0.01 & 0.02 & -0.02 & -0.71 \\ -0.01 & 0.02 & -0.02 & -0.71  
\end{array}\right] \ , 
\end{eqnarray}

\bigskip

\noindent
{\bf Acknowledgments:}  
G.B. acknowledges support from the John Simon Guggenheim Foundation, and the Conselho Nacional de Desenvolvimento Cientifico e Tecnologico (CNPq). 
Fermilab is operated by Fermi 
Research Alliance, LLC under contract no. DE-AC02-07CH11359 with the United States Department of Energy. L.D. 
thanks Alejandro Szynkman for very useful discussions on four body decays and Daniel de Flori{\'a}n for suggesting the use of CompHEP for many-body decays.
R.D.M. acknowledges support from FAPESP (Sao Paulo State Research Foundation).


\end{document}